\documentclass[showpacs,showkeys,preprintnumbers,amssymb,aps,superscriptaddress,twocolumn]{revtex4-2}
\usepackage{bm}
\usepackage{tabularray}
\usepackage{amsmath,graphicx,array,float,amsthm,amssymb}
\usepackage{tikz,color,fullpage,epsf,multirow}
\usepackage[colorlinks={true}]{hyperref}
\hypersetup{colorlinks=true,linkcolor=blue,citecolor=blue,urlcolor=blue}
\usepackage{amsmath,graphicx,verbatimbox,array,float}
\usepackage{indentfirst,orcidlink}

\makeatletter
\renewcommand{\p@subsection}{}
\makeatother
\usepackage{braket}
\usepackage{pstricks}
\usepackage{diagbox}
\usepackage{amssymb}
\usepackage[percent]{overpic}
\usepackage{bbold,soul}

\begin{document}
\title{Towards Distributed Quantum Computing: Quantum Gate Teleportation via Satellite-Distributed Entanglement under Atmospheric Depolarization}

\author{Artur Czerwinski~\!\!\orcidlink{0000-0003-0625-8339}}\email{aczerwin@umk.pl}
\affiliation{Institute of Physics, Faculty of Physics, Astronomy and Informatics, Nicolaus Copernicus University in Torun, ul. Grudziadzka 5, 87-100 Torun, Poland}

\author{Jakub J. Borkowski~\!\!\orcidlink{0009-0007-1526-9394}}\email{jborkowski@doktorant.umk.pl}
\affiliation{Institute of Physics, Faculty of Physics, Astronomy and Informatics, Nicolaus Copernicus University in Torun, ul. Grudziadzka 5, 87-100 Torun, Poland}

\author{Maciej Wiśniewski~\!\!\orcidlink{0009-0008-1262-5186}}\email{u55121@sdr.wat.edu.pl}
\affiliation{Institute of Optoelectronics, Military University of Technology, ul. gen. Sylwestra Kaliskiego 2, 00-908 Warsaw, Poland}

\author{Saeed Haddadi~\!\!\orcidlink{0000-0002-1596-0763}}\email{haddadi@ipm.ir}
\affiliation{School of Particles and Accelerators, Institute for Research in Fundamental Sciences (IPM), P.O. Box 19395-5531, Tehran, Iran}

\begin{abstract}
We investigate quantum gate teleportation as a primitive for satellite-enabled distributed quantum computing. In the considered architecture, a satellite distributes entangled photon pairs to two optical ground stations, which transfer the shared quantum resource to remote quantum processing units. This entanglement is then used to implement a nonlocal controlled-NOT gate between two circuit qubits located in different processors. We formulate the minimal four-qubit protocol, consisting of two network qubits that provide the entangled resource and two circuit qubits on which the distributed gate is performed. Atmospheric propagation effects in the gravitational field are modeled through an effective two-qubit depolarizing channel whose parameters depend on atmospheric and optical quantities, including the Fried parameter and receiver aperture. The performance of the protocol is quantified by the fidelity between the ideal CNOT output state and the state produced by noisy quantum gate teleportation. Our results show that the fidelity is governed by the quality of the distributed entanglement and depends jointly on atmospheric coherence, aperture size, and effective depolarization strength. These findings connect physical-layer modeling of satellite optical channels with gate-level performance in distributed quantum computing.
\end{abstract}


\maketitle

\section{Introduction}

Quantum computing uses the principles of quantum mechanics to process information in ways that are essentially inaccessible to classical computers.  
Since the fundamental proposals of quantum algorithms and fault-tolerant architectures, quantum computing has become one of the most active research fields in modern science and technology~\cite{Nielsen2000,shor1997polynomial,grover1997quantum,ladd2010quantum}. The operating basis of quantum computing is the quantum circuit model, in which information is manipulated through sequences of quantum logic gates that operate on one or more qubits. These gates are unitary transformations that preserve quantum coherence and allow arbitrary quantum algorithms to be decomposed into elementary operations capable of approximating any quantum computation with arbitrary accuracy. Hence, the realization of high-fidelity quantum gates is one of the main requirements for scalable quantum information processing \cite{Nielsen2000,barenco1995elementary,divincenzo2000physical}.

Despite significant advances in the development of quantum hardware platforms, including superconducting circuits, trapped ions, photonic systems, neutral atoms, and semiconductor qubits, the size and complexity of individual quantum processors are still limited by technological constraints.
These include finite coherence times, fabrication challenges, control errors, and qubit coupling limitations. Hence, increasing attention has been directed towards distributed quantum computing (DQC), in which multiple smaller quantum processors are interconnected via quantum communication channels and collectively perform computational tasks that exceed the capabilities of any single device. In such architectures, the quantum network becomes an integral component of the computing infrastructure, rather than just a communication layer~\cite{cirac1999distributed,wehner2018quantum,cuomo2020towards,covey2023networks}.
A key challenge in DQC is implementing nonlocal quantum gates between qubits on different processors~\cite{main2025distributed,sarvaghad2021distributed,gyongyosi2021distributed}. Direct physical interactions are usually not available when computing nodes are separated by large distances. However, quantum gate teleportation (QGT) provides an elegant solution to this problem by transforming pre-shared entanglement and classical communication into an effective nonlocal gate operation \cite{Gottesman1999,Eisert2000,LEUNG2004}. In this way, entanglement serves as a consumable resource that enables remote quantum processing while maintaining the modular structure of the network~\cite{monroe2014modular}. Thus, QGT has emerged as a fundamental principle for fault-tolerant quantum computing, distributed quantum algorithms, quantum repeater architectures, and measurement-based quantum computing~\cite{gottesman1999demonstrating,eisert2000optimal,zhou2000methodology}.

The core idea of QGT is that a quantum logic gate can be implemented between two distant parties without requiring a direct physical interaction between the corresponding qubits. Instead, the nonlocal operation is realized by combining a pre-shared entangled state with local operations, measurements, classical communication, and feed-forward corrections~\cite{gottesman1999demonstrating}. This feature makes QGT particularly relevant for distributed quantum computing, where entangling gates must be performed between quantum processors located in different modules or network nodes. In such architectures, direct coupling between circuit qubits may be technically infeasible, especially when the nodes are connected through free-space optical, space-to-ground, or inter-satellite links \cite{kaushal2016optical}. The feasibility of this approach therefore depends on the ability to distribute high-quality entanglement over long distances. Satellite-based quantum links provide a promising route to this goal because they can reduce the distance-dependent losses that strongly limit direct fiber-based transmission~\cite{Bedington2017,Khatri2021}. This possibility has already been demonstrated experimentally, most notably by the Micius satellite mission, which achieved distribution of entangled photon pairs to two ground stations separated by \(1203~\mathrm{km}\)~\cite{Yin2017}. These results show that satellite-mediated entanglement distribution can provide the physical resource required for higher-level quantum network protocols. In this context, integrating satellite-distributed entanglement with QGT represents a natural step towards distributed quantum computing over global-scale quantum networks~\cite{deForges2023}.

From a broader perspective, the convergence of quantum computing and quantum networking is expected to define the next generation of quantum technologies. Future quantum infrastructures are likely to include geographically distributed processors, interconnected via terrestrial and space-based quantum channels, forming a quantum internet capable of supporting secure communications, distributed sensing, and large-scale quantum computing. In this view, the ability to implement quantum gates remotely with high precision becomes as important as the ability to distribute entanglement itself. Therefore, understanding the performance of QGT under realistic communication conditions is essential to assess the feasibility of large-scale networked quantum computing systems. This motivation is particularly necessary in the context of satellite-assisted quantum networks, where global-scale entanglement distribution may eventually enable the realization of planetary-scale distributed quantum processors~\cite{wehner2018quantum,kimble2008quantum,simon2017towards,pirandola2021limits,pirandola2021satellite}.

\begin{figure}[t]
    \centering
    \includegraphics[width=0.99\columnwidth]{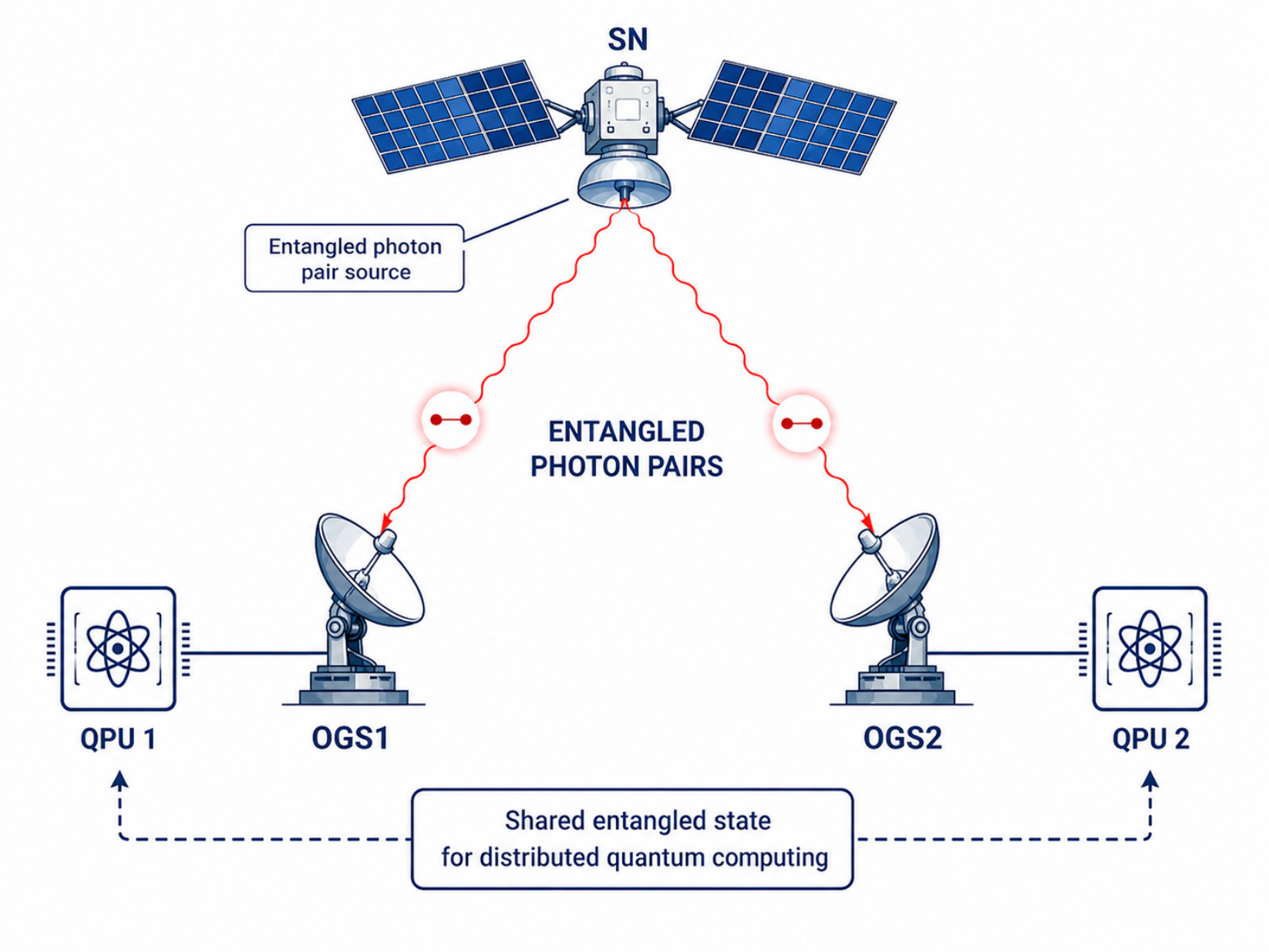}
    \caption{Conceptual illustration of satellite-enabled entanglement distribution for distributed quantum computing. A satellite node (SN) hosts a source of entangled photon pairs and distributes the photons through two optical downlinks to optical ground stations denoted by OGS1 and OGS2. The received quantum states are subsequently transferred through fiber-based infrastructure to quantum processing units (QPUs), thereby establishing a shared entangled state between remote processors.}
    \label{figure00g1}
\end{figure}

In the present work, we focus on QGT not only as a standalone application, but also as a building block for DQC. The vision of DQC relies on linking together multiple spatially separated quantum processors to act as a single, larger quantum computer. Achieving this goal requires the repeated execution of non-local quantum gates between modules located at different sites. In practice, this would mean that many entangled photon pairs need to be reliably distributed and maintained across the nodes of the network. Once this entanglement is established, series of QGT operations could be performed to implement the logical circuit of a distributed algorithm.
 
Satellite links can provide a physical layer for distributing quantum resources between distant quantum processing units. 
In the scheme shown in Fig.~\ref{figure00g1}, a satellite node (SN) hosts a source of entangled photon pairs and distributes the two photons through optical downlinks to two optical ground stations---OGS1 and OGS2. After coupling the received photonic states to local quantum processing units, QPU1 and QPU2, share an entangled state that can be used as a nonlocal quantum resource for DQC protocols (or other applications).

The ability to teleport gates across space-based links would therefore play a decisive role in scaling up quantum computers beyond the size of individual physical devices. By investigating the performance of QGT under realistic atmospheric and channel conditions, our work provides a first step towards assessing the viability of such distributed architectures in the space domain.

The paper is organized as follows. Section~\ref{depoltheory} introduces the theoretical framework for modeling effective depolarization in atmospheric channels. Section~\ref{effectivedepolarization} applies this framework to uplink and downlink transmission, examining how atmospheric and optical parameters, including the Fried parameter, wind speed, and aperture size, influence the depolarization probability. Section~\ref{remotecnotgate} investigates the impact of effective depolarization on QGT and comprises four parts. Section~\ref{dualchannel} formulates the dual-link channel used to distribute entangled photon pairs from a satellite to two OGSs. Section~\ref{remotecnotgateth} then presents the QGT protocol, followed by an analysis of its fidelity and the physical interpretation of the results in Section~\ref{remotecnotgateres}. Section~\ref{sec:DifferentCircuit} examines how the input-state geometry of the circuit qubits affects QGT fidelity. Finally, Section~\ref{conclusions} summarizes the main findings and conclusions.

\section{Mathematical Framework for Effective Atmospheric Depolarization}\label{depoltheory}

We describe effective noise in the combined atmospheric link and optical receiver using a depolarizing channel. The physical mechanism considered here is the conversion of turbulence-induced
spatial disturbances into polarization fluctuations through spatially dependent receiver transformations, followed by averaging over unresolved spatial modes or temporal realizations. Scalar
refractive-index fluctuations alone need not change the polarization of an initially separable spatial--polarization state. Accordingly, the model requires both atmospheric spatial fluctuations and a
mechanism coupling these fluctuations to polarization.

The spatial-coherence formulation of Wang et al.~\cite{Wang2011} provides a starting point for describing atmospheric averaging. However, obtaining a single-qubit depolarizing channel additionally requires a receiver model and an assumption concerning the symmetry of the effective polarization noise. In this section, we derive an exponential depolarization law using Kolmogorov phase statistics and an isotropic
diffusion approximation for the residual polarization transformations.

\subsection{Atmospheric Coherence and the Kolmogorov Limit}

Wang et al.~\cite{Wang2011} describe how atmospheric fluctuations affect the coherence between field contributions originating at two transverse positions, $\mathbf r_1$ and $\mathbf r_2$, in the source
plane. Both contributions propagate to the same transverse position $\mathbf r$ in the receiver plane at propagation distance $z$. For fixed $\mathbf r$ and $z$, the atmospheric factor in the propagation integral is
\begin{equation}
\label{eq:wang_atmospheric_kernel}
B(\mathbf r_1,\mathbf r_2)
=
\left\langle
\exp\!\left[
\psi^*(\mathbf r,\mathbf r_1,z)
+
\psi(\mathbf r,\mathbf r_2,z)
\right]
\right\rangle_m,
\end{equation}
where $\psi(\mathbf r,\mathbf r_j,z)$ is the random complex perturbation accumulated during propagation from $\mathbf r_j$ to $\mathbf r$. Its real and imaginary parts describe the log-amplitude and phase fluctuations, respectively, and $\langle\cdot\rangle_m$ denotes the average over realizations of the turbulent medium.

Using a quadratic approximation to the wave structure function, Wang et al. obtain [Eq.~(13) of Ref.~\cite{Wang2011}]
\begin{equation}
B(\mathbf r_1,\mathbf r_2)
\simeq
\exp\!\left(-\frac{s^2}{\rho_0^2}\right),
\qquad
s=|\mathbf r_1-\mathbf r_2|.
\end{equation}
Here $\rho_0$ is the transverse coherence length associated with atmospheric propagation in this approximation. It sets the separation between source points at which the atmospheric factor decreases to $1/e$. The Gaussian dependence on $s$ allows the propagation integrals to be evaluated analytically.

To obtain the Kolmogorov scaling, we consider a normalized phase-only coherence factor at the receiver pupil. Amplitude fluctuations and transmission losses are treated separately in this reduction. For zero-mean Gaussian phase increments,
\begin{equation}
\label{eq:phase_coherence_structure}
B_\phi(s)
=
\left\langle
e^{i[\phi(\mathbf r_2)-\phi(\mathbf r_1)]}
\right\rangle
=
\exp\!\left[-\frac{1}{2}D_\phi(s)\right],
\end{equation}
and
\begin{equation}
D_\phi(s)
=
\left\langle
[\phi(\mathbf r_2)-\phi(\mathbf r_1)]^2
\right\rangle,
\end{equation}
is the phase structure function.

The transverse scaling can be understood from a phase spectrum $\Phi_\phi(Q)\propto Q^{-\alpha}$ in the inertial range. For statistically homogeneous and isotropic fluctuations,
\begin{equation}
\begin{aligned}
D_\phi(s)
&\propto
\int_0^\infty
Q^{1-\alpha}\,[1-J_0(Qs)]\,dQ
\\
&=
s^{\alpha-2}
\int_0^\infty
u^{1-\alpha}\,[1-J_0(u)]\,du,
\end{aligned}
\end{equation}
where $Q$ is the transverse spatial wavenumber and $J_0$ is the Bessel function of the first kind. The idealized integral converges for $2<\alpha<4$. For Kolmogorov turbulence, $\alpha=11/3$, giving $D_\phi(s)\propto s^{5/3}$.

With the conventional Fried-parameter normalization, the results are
\cite{Fried1966,Andrews2005}
\begin{equation}
\label{eq:kolmogorov_phase_structure}
D_\phi(s)
=
6.88\left(\frac{s}{r_0}\right)^{5/3}
\end{equation}
and
\begin{equation}
B_\phi(s)
=
\exp\!\left[
-3.44\left(\frac{s}{r_0}\right)^{5/3}
\right],
\end{equation}
where $r_0$ denotes the coherence length (Fried parameter). These expressions apply to separations in the inertial range, between the inner and outer turbulence scales, with $r_0$ defined for the relevant propagation geometry and reference plane.

Eq.~\eqref{eq:kolmogorov_phase_structure} describes spatial coherence, not atmospheric depolarization. In particular, $1-B_\phi(s)$ cannot generally be identified with a polarization error probability. Moreover, replacing the non-Kolmogorov spectral exponent by $11/3$ in the quadratic approximation of Wang et al. does not itself recover the $s^{5/3}$ dependence: the inertial-range structure function must be retained before
performing that approximation.

\subsection{Receiver-weighted Measure of Spatial Disturbance}

Let $W(\mathbf r)\geq 0$ denote a normalized effective spatial weight at the receiver pupil,
\begin{equation}
\int W(\mathbf r)\,d^2\mathbf r=1.
\end{equation}
For a fixed receiver-mode shape, write
\begin{equation}
W(\mathbf r)
=
a_{\mathrm{eff}}^{-2}
w(\mathbf r/a_{\mathrm{eff}}),
\qquad
\int w(\mathbf x)\,d^2\mathbf x=1,
\end{equation}
where $a_{\mathrm{eff}}$ is the characteristic transverse scale. The quantity $a_{\mathrm{eff}}$ must refer to the same optical plane as $r_0$ from Eq.~(\ref{eq:kolmogorov_phase_structure}); a mode size specified after relay optics must therefore be referred back to the pupil.

We define a dimensionless spatial-disturbance parameter by
\begin{equation}
\label{eq:receiver_disturbance}
\chi
=
\frac{1}{2}
\iint
W(\mathbf r_1)W(\mathbf r_2)
D_\phi(|\mathbf r_1-\mathbf r_2|)
\,d^2\mathbf r_1\,d^2\mathbf r_2.
\end{equation}
Using Eq.~\eqref{eq:kolmogorov_phase_structure} and rescaling the coordinates gives
\begin{equation}
\label{eq:chi_kolmogorov}
\chi
=
3.44\,C_w
\left(\frac{a_{\mathrm{eff}}}{r_0}\right)^{5/3},
\end{equation}
where
\begin{equation}
C_w
=
\iint
w(\mathbf x)w(\mathbf y)
|\mathbf x-\mathbf y|^{5/3}
\,d^2\mathbf x\,d^2\mathbf y,
\end{equation}
is a dimensionless geometrical factor, assumed finite.

This definition extracts the Kolmogorov scaling of a receiver-weighted phase variance. It does not assert that the aperture-averaged coherence equals $\exp(-\chi)$; in general, averaging an exponential is not equivalent to exponentiating its average.

\subsection{Conversion into Effective Polarization Noise}

It is important to distinguish between direct depolarization and effective depolarization induced by unresolved degrees of freedom. In clear air, atmospheric turbulence is primarily a scalar refractive-index fluctuation. It mainly distorts the phase and amplitude of the optical field and does not necessarily act as a strongly birefringent medium. Therefore, a polarization qubit may remain nearly pure in a single realization of the optical channel.

However, the polarization state detected at the receiver may still become mixed if polarization becomes correlated with the spatial degree of freedom. For example, an initially separable pure state $\ket{\psi}_{\mathrm{pol}}$ may evolve into
\begin{equation}
\ket{\Psi}
=
\int
u(\mathbf r)\ket{\mathbf r}
\otimes U(\mathbf r)\ket{\psi}_{\mathrm{pol}}
\,d^2\mathbf r,
\end{equation}
where $u(\mathbf r)$ denotes the spatial mode at the receiver and $U(\mathbf r)$ represents a local effective polarization transformation associated with position, local angle of incidence, or receiver coupling conditions.

If $U(\mathbf r)$ is approximately constant across the detected mode, the output polarization state remains pure up to a unitary transformation. If $U(\mathbf r)$ varies significantly over the detected spatial region, then the polarization degree of freedom becomes correlated with the spatial degree of freedom. Since the detector typically does not resolve the full spatial structure, the observed polarization state is obtained by tracing out the spatial mode:
\begin{equation}
\rho_{\mathrm{pol,out}}= \mathrm{Tr}_{\mathrm{space}}
\left[\ket{\Psi}\!\bra{\Psi}\right],
\end{equation}
which can be rewritten as
\begin{equation}
\label{eq:spatial_random_unitary}
\rho_{\mathrm{pol,out}}
=
\int
|u(\mathbf r)|^2
U(\mathbf r)\rho_{\mathrm{pol,in}}U^\dagger(\mathbf r)
\,d^2\mathbf r,
\end{equation}
with $\int |u(\mathbf r)|^2\,d^2\mathbf r=1$. Temporal averaging can produce an analogous random-unitary map.

After compensating for a deterministic polarization rotation,
we model the residual polarization transformation as a sequence
of weak, independent, isotropic random rotations. We use the
receiver-weighted phase variance $\chi$, defined in
Eq.~\eqref{eq:receiver_disturbance}, to parameterize this
effective diffusion process. We assume that an increase $d\chi$
in the spatial phase variance contributes a polarization-rotation
variance $g\,d\chi$ about each axis, where $g\geq 0$ is a
dimensionless coupling coefficient specific to the optical system.

The unitary associated with an infinitesimal rotation is
\begin{equation}
U_{\delta}=\exp\!\left[-\frac{i}{2}\sum_{j=x,y,z}\delta\theta_j\sigma_j
\right],
\end{equation}
where the angular increments satisfy
\begin{equation}
\label{eq:rotation_covariance}
\mathbb E[\delta\theta_j]=0,
\qquad
\mathbb E[\delta\theta_j\delta\theta_k]=g\,\delta_{jk}\,d\chi.
\end{equation}
For constant $g$, the accumulated variance of the rotation
increments about each axis is therefore $g\chi$. For a given
link and receiver, Eq.~\eqref{eq:chi_kolmogorov} specifies the
value of $\chi$ at which the effective channel is evaluated.

Expanding $U_{\delta}\rho U_{\delta}^{\dagger}$ to second order and averaging yields
\begin{equation}
\label{eq:polarization_diffusion}
\frac{d\rho}{d\chi}
=
-\frac{g}{8}
\sum_{j=x,y,z}
[\sigma_j,[\sigma_j,\rho]].
\end{equation}
Using
\begin{equation}
\sum_{j=x,y,z}\sigma_j\rho\sigma_j
=
2\mathbb I_2-\rho,
\end{equation}
for a normalized qubit state, we obtain
\begin{equation}
\frac{d\rho}{d\chi}
=
-g\left(\rho-\frac{\mathbb I_2}{2}\right).
\end{equation}
With $\rho(0)=\rho_{\mathrm{in}}$, the solution is
\begin{equation}
\label{eq:diffusion_depolarizing_solution}
\rho(\chi)
=
e^{-g\chi}\rho_{\mathrm{in}}
+
\left(1-e^{-g\chi}\right)\frac{\mathbb I_2}{2}.
\end{equation}
Thus, isotropic polarization diffusion generates precisely a single-qubit depolarizing channel.

Combining Eqs.~\eqref{eq:chi_kolmogorov} and \eqref{eq:diffusion_depolarizing_solution} gives
\begin{equation}
\label{eq:pdep_kolmogorov}
p_{\mathrm{dep}}
=
1-\exp\!\left[
-\kappa
\left(\frac{a_{\mathrm{eff}}}{r_0}\right)^{5/3}
\right],
\end{equation}
where
\begin{equation}
    \kappa=3.44\,g\,C_w.
\end{equation}
The exponent $5/3$ in Eq.~(\ref{eq:pdep_kolmogorov}) follows from the Kolmogorov phase structure
function, whereas the exponential contraction follows from the isotropic diffusion approximation. The coefficient $\kappa$ combines the receiver geometry and its sensitivity to spatial--polarization coupling. It is not fixed by atmospheric statistics alone.

\subsection{Relation to the Degree of Polarization}

For a detected single photon, we write the polarization state as
\begin{equation}
\rho
=
\frac{1}{2}
\left(\mathbb I_2+ \boldsymbol{ \mathcal{S}}\cdot\boldsymbol\sigma\right),
\end{equation}
where the Bloch-vector length $P_1=| \boldsymbol{\mathcal{S}}|$ gives the
first-order degree of polarization. The depolarizing channel
in Eq.~\eqref{eq:diffusion_depolarizing_solution} contracts the
Bloch vector by the same factor along each axis,
\begin{equation}
\mathbf \mathcal{S}_{\mathrm{out}}
=
e^{-g\chi} \boldsymbol{\mathcal{S}}_{\mathrm{in}}
=
(1-p_{\mathrm{dep}}) \boldsymbol{\mathcal{S}}_{\mathrm{in}}.
\end{equation}
Thus, $1-p_{\mathrm{dep}}$ is the fraction of the initial
polarization retained by the channel. For a pure input state,
$P_{1,\mathrm{in}}=1$ and
$P_{1,\mathrm{out}}=1-p_{\mathrm{dep}}$.

Wang et al.~\cite{Wang2011} use the quantum degree of polarization
\begin{equation}
P_{\mathrm W}
=
\frac{|\langle\hat{\boldsymbol{\mathcal{S}}}\rangle|}
{\sqrt{\langle\hat S_0(\hat S_0+2)\rangle}}.
\end{equation}
In the single-photon sector, $\hat S_0=1$ and
$|\langle\hat{\mathbf S}\rangle|=|\boldsymbol{\mathcal{S}}|$, giving
$P_{\mathrm W}=P_1/\sqrt{3}$. Consequently, even a pure
single-photon polarization state has $P_{\mathrm W}=1/\sqrt{3}$
under this normalization. For input and output states conditioned
on one photon, the normalization factor cancels in the ratio,
\begin{equation}
\frac{P_{\mathrm W,out}}{P_{\mathrm W,in}}
=
\frac{P_{1,\mathrm{out}}}{P_{1,\mathrm{in}}}
=
1-p_{\mathrm{dep}},
\qquad P_{1,\mathrm{in}}>0.
\end{equation}

\subsection{Generalized Model and Validity}

For this subsection, we retain the generalized expression
\begin{equation}
\label{eq:pdep_general}
p_{\mathrm{dep}}
=
1-\exp\!\left[
-\kappa
\left(
\frac{a_{\mathrm{eff}}}{r_0}
\right)^\epsilon
\right],
\end{equation}
where $\kappa\geq 0$ is a system-dependent coupling coefficient and $\epsilon>0$ is an effective scaling exponent. The derivation above gives $\epsilon=5/3$ when the relevant separations lie in the Kolmogorov inertial range, the normalized receiver-mode shape is fixed, and the coupling coefficient $g$ is approximately independent of turbulence strength. Departures from these assumptions can require a fitted exponent or a more detailed channel model.

Setting $x=a_{\mathrm{eff}}/r_0$, the weak-disturbance expansion is
\begin{equation}
p_{\mathrm{dep}}
=
\kappa x^\epsilon
-\frac{\kappa^2}{2}x^{2\epsilon}
+O(x^{3\epsilon}),
\qquad x\to0.
\end{equation}
For fixed $\kappa>0$, the effective model also satisfies
\begin{equation}
\lim_{x\to\infty}p_{\mathrm{dep}}=1.
\end{equation}
This saturation is a property of the isotropic diffusion model, rather than a universal prediction for atmospheric propagation. A polarization-independent optical system has $g=\kappa=0$ and therefore no depolarization through the mechanism modeled here, even when spatial coherence is strongly degraded.

The ratio $a_{\mathrm{eff}}/r_0$ measures the transverse extent sampled by the receiver relative to the atmospheric coherence length; it is not itself a count of coherence cells. For an aperture-limited receiver, $a_{\mathrm{eff}}\simeq D/2$ can provide a practical scale estimate. Its use remains
conditional on the actual mode weighting and spatial--polarization coupling of the receiver.

Consequently, Eq.~\eqref{eq:pdep_general} should be regarded as a Kolmogorov-motivated effective channel model, connected to the coherence-averaging framework of Wang et al., with an explicit additional assumption of isotropic polarization diffusion. Calibration or end-to-end optical simulation is required to determine $\kappa$ and establish the range over which this reduction accurately describes the detected polarization state.

\section{Practical modeling of effective atmospheric depolarization}\label{effectivedepolarization}

\subsection{Models for the Fried Parameter $r_0$}

The Fried parameter $r_0$ quantifies the transverse coherence length of the turbulent wavefront. It is a standard parameter in the theory of optical propagation through atmospheric turbulence and determines the characteristic transverse distance over which phase distortions remain correlated. For monochromatic light with wavenumber $k=2\pi/\lambda$, the following forms are commonly used \cite{Fried1966,Fante1975,Andrews2005}.

\paragraph{Constant $C_n^2$.}
If the refractive index structure parameter $C_n^2$ is assumed constant along the propagation path of length $L$, the Fried parameter is
\begin{equation}
\label{eq:r0_const}
r_0^{\mathrm{(const)}}
=
\left(
0.423\,k^2\,C_n^2\,L
\right)^{-3/5}.
\end{equation}
This expression is useful for simplified horizontal links or for approximate estimates in which the turbulence strength is represented by a single effective value of $C_n^2$.

\paragraph{Altitude-dependent $C_n^2(h)$ and downlink transmission.}
For an inhomogeneous atmosphere, where the refractive index structure parameter varies with altitude $h$, the Fried parameter for downlink propagation can be written as
\begin{equation}
\label{eq:r0_integral}
r_0^{\mathrm{(down)}}
=
\left[
0.423\,k^2 \sec \zeta
\int_{H_{\mathrm{OGS}}}^{H}
C_n^2(h)\,dh
\right]^{-3/5},
\end{equation}
where $\zeta$ is the zenith angle, $H_{\mathrm{OGS}}$ is the altitude of the OGS above sea level, and $H$ denotes the maximum altitude of the turbulent atmospheric layer. The factor $\sec \zeta$ accounts for the increased optical path length in slant propagation relative to vertical propagation.

For satellite optical links, the Hufnagel--Valley (H--V) model is frequently used to estimate the altitude-dependent behavior of the refractive index structure parameter \cite{hufnagel1964modulation,valley1980isoplanatic}:
\begin{equation}
\begin{aligned}
\label{hvmodel}
C_n^2(h)
=&
8.148\times 10^{-56}\,v_{\mathrm{rms}}^2\,h^{10}e^{-h/1000}
\\
&+
2.7\times 10^{-16}e^{-h/1500}
\\
&+
C_0\,e^{-H_{\mathrm{OGS}}/700}
e^{-(h-H_{\mathrm{OGS}})/100}.
\end{aligned}
\end{equation}
Here, $h$ and $H_{\mathrm{OGS}}$ are expressed in meters, $v_{\mathrm{rms}}$ is the root-mean-square wind speed, and $C_0$ characterizes the near-ground turbulence strength (e.g. $C_0 = 1.7 \times 10^{-14}$~m$^{-2/3}$ as a typical nighttime value). The final exponential term is written relative to the OGS altitude, so that near-ground turbulence decays with height above the local station.

The wind speed $v_{\mathrm{rms}}$ can be estimated using the Bufton wind model \cite{bufton1973comparison}. Different values of $v_{\mathrm{rms}}$ are used to represent different atmospheric conditions. For example, $v_{\mathrm{rms}}=21$~m/s is often used for weak-wind scenarios, while $v_{\mathrm{rms}}=57$~m/s corresponds to strong-wind conditions \cite{calvo2014transmitter,Czerwinski2026}. Intermediate values, such as $v_{\mathrm{rms}}=26.25$~m/s, can be used to represent moderate wind conditions \cite{Czerwinski2025}. Such a choice avoids restricting the analysis to either the most favorable or the most challenging atmospheric regimes.

\paragraph{Altitude-dependent $C_n^2(h)$ and uplink transmission.}

\begin{figure}[t]
    \centering
    \includegraphics[width=0.99\columnwidth]{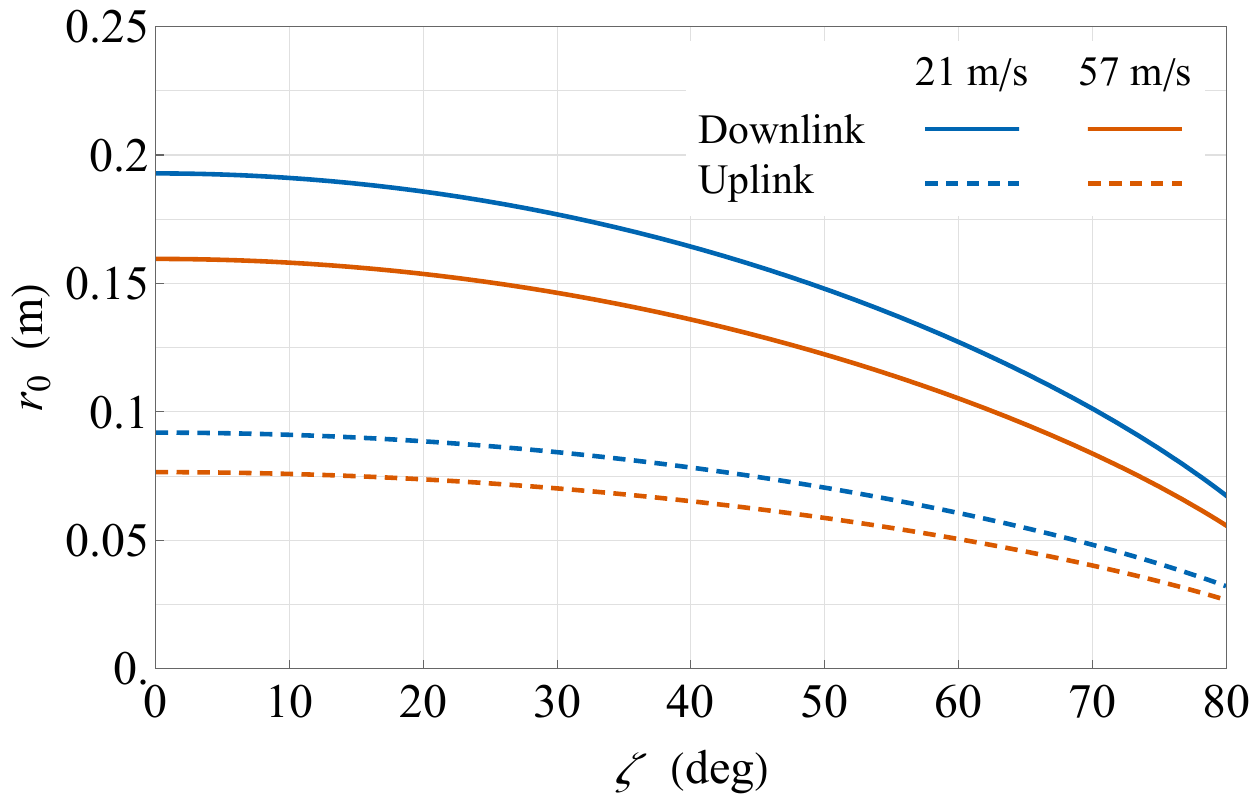}
    \caption{The Fried parameter for downlink transmission (solid curves) and uplink transmission (dashed curves) as a function of the zenith angle, shown for two representative wind speeds, $v_{\mathrm{rms}}=21$~m/s and $v_{\mathrm{rms}}=57$~m/s, corresponding to weak and strong turbulence conditions, respectively.}
    \label{figure03}
\end{figure}

For uplink propagation, the optical beam originates at the ground and expands as a spherical wave while traversing turbulence at increasing altitudes. In this case, turbulence encountered close to the transmitter has a particularly strong influence on the resulting wavefront distortion. Under the spherical-wave approximation, the Fried parameter for uplink transmission can be written as \cite{Fante1975}
{\small\begin{equation}
\label{eq:r0_uplink}
r_0^{\mathrm{(up)}}
=
\left[
1.46\,k^2 \sec \zeta
\int_{H_{\mathrm{OGS}}}^{H}
C_n^2(h)
\left(1-\xi\right)^{5/3}
dh
\right]^{-3/5},
\end{equation}}
where
\begin{equation}
\label{eq:xi_uplink}
\xi
=
\frac{h-H_{\mathrm{OGS}}}{H-H_{\mathrm{OGS}}}.
\end{equation}
The weighting factor $\left(1-\xi\right)^{5/3}$ reflects the altitude-dependent contribution of turbulence to the coherence properties of the uplink spherical wave. As in the downlink case, $k$ is the optical wavenumber, $\zeta$ is the zenith angle, and the integral extends over the turbulent atmospheric layer from the OGS altitude $H_{\mathrm{OGS}}$ up to altitude $H$.

Equations~\eqref{eq:r0_const}, \eqref{eq:r0_integral}, and \eqref{eq:r0_uplink} summarize the Fried parameter for three levels of atmospheric modeling. Equation~\eqref{eq:r0_const} applies to a homogeneous atmosphere with constant $C_n^2$, while Eqs.~\eqref{eq:r0_integral} and \eqref{eq:r0_uplink} describe altitude-dependent turbulence profiles for downlink and uplink transmission, respectively.

The results shown in Fig.~\ref{figure03} highlight the difference between downlink and uplink propagation. For downlink transmission, the Fried parameter typically remains larger because the beam reaches the strongly turbulent lower atmosphere after propagating through the less turbulent upper layers. In contrast, the uplink configuration exhibits smaller coherence lengths because the beam encounters strong near-ground turbulence immediately after transmission, when the beam is still relatively narrow. This makes the uplink more sensitive to low-altitude turbulence. In both geometries, the Fried parameter decreases with increasing zenith angle, reflecting the longer slant path through the turbulent atmosphere.

\subsection{Effective Transverse Scale $a_{\mathrm{eff}}$ and Relation to Aperture $D$}

The effective transverse scale $a_{\mathrm{eff}}$ represents the spatial scale over which the polarization state is effectively averaged at the receiver. Depending on the detection architecture, it may correspond to the spot radius at the detector plane, the mode radius at the fiber input, or an effective aperture radius after relay optics. For an aperture-limited receiving telescope, a practical engineering approximation allows us to take by definition
\begin{equation}
\label{eq:w0_vs_D}
a_{\mathrm{eff}} \equiv \frac{D}{2},
\end{equation}
where $D$ is the telescope diameter.

This approximation should not be interpreted as a fundamental law. It is a convenient way to relate the spatial averaging scale in the phenomenological depolarization model to the physical size of the receiver aperture. If the receiver is directly aperture-limited and does not resolve spatial modes, a larger aperture collects a wider portion of the turbulent wavefront and can therefore sample more mutually less coherent regions. In such a case, increasing $D$ increases the ratio $a_{\mathrm{eff}}/r_0$ and may increase the effective depolarization probability.

When a single-mode fiber is used, the situation is more subtle. The relevant value of $a_{\mathrm{eff}}$ should correspond to the optical mode that is actually coupled into the fiber, after accounting for relay optics, focusing, truncation, and spatial filtering. In this configuration, the telescope aperture affects both the collected power and the mode-matching conditions, but the effective polarization averaging scale may differ from the simple radius $D/2$. Therefore, Eq.~\eqref{eq:w0_vs_D} should be regarded as a first-order approximation appropriate for aperture-based estimates, while more accurate simulations should compute $a_{\mathrm{eff}}$ from the full optical receiver model.

Combining Eqs.~\eqref{eq:pdep_general} and \eqref{eq:w0_vs_D} shows that the effective depolarization probability increases with the ratio $D/r_0$ in an aperture-limited model. This does not mean that a larger telescope intrinsically depolarizes photons. Rather, it means that a larger receiver may collect light from more spatially diverse portions of the wavefront. If these portions are associated with different effective polarization transformations, and if the receiver does not resolve these spatial contributions, the detected polarization state becomes an incoherent mixture. Conversely, a larger Fried parameter corresponds to larger atmospheric coherence cells and therefore reduces the number of statistically distinct spatial contributions sampled by the receiver. This leads to a lower effective depolarization probability.

\subsection{Probability of Depolarization: Examples and Discussion}

\begin{figure*}[t]
    \centering
    \begin{overpic}[height=3.99cm,keepaspectratio]{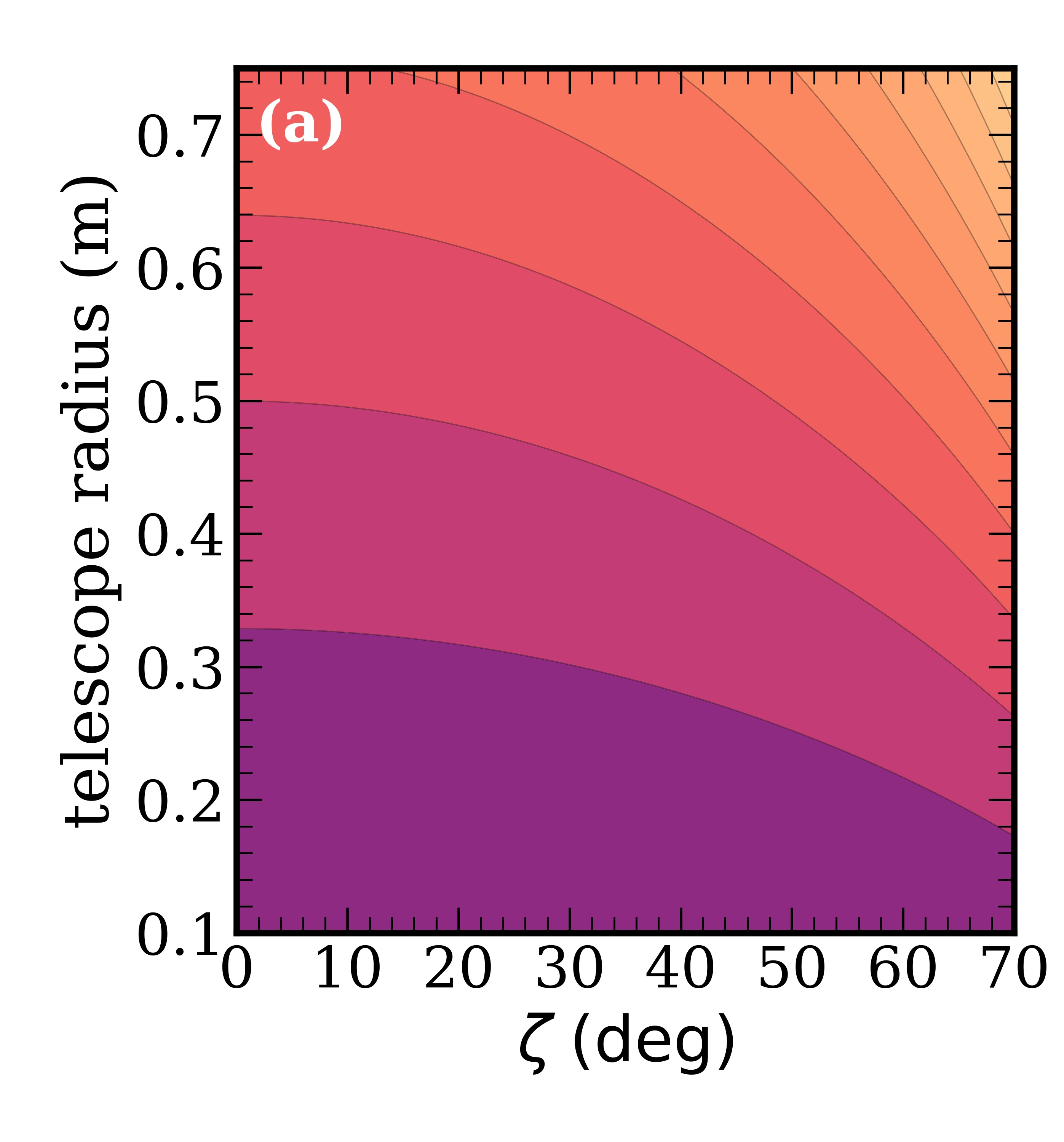}
        \put(4,91){\white\textbf{(a)}}
    \end{overpic}
    \begin{overpic}[height=3.99cm,keepaspectratio]{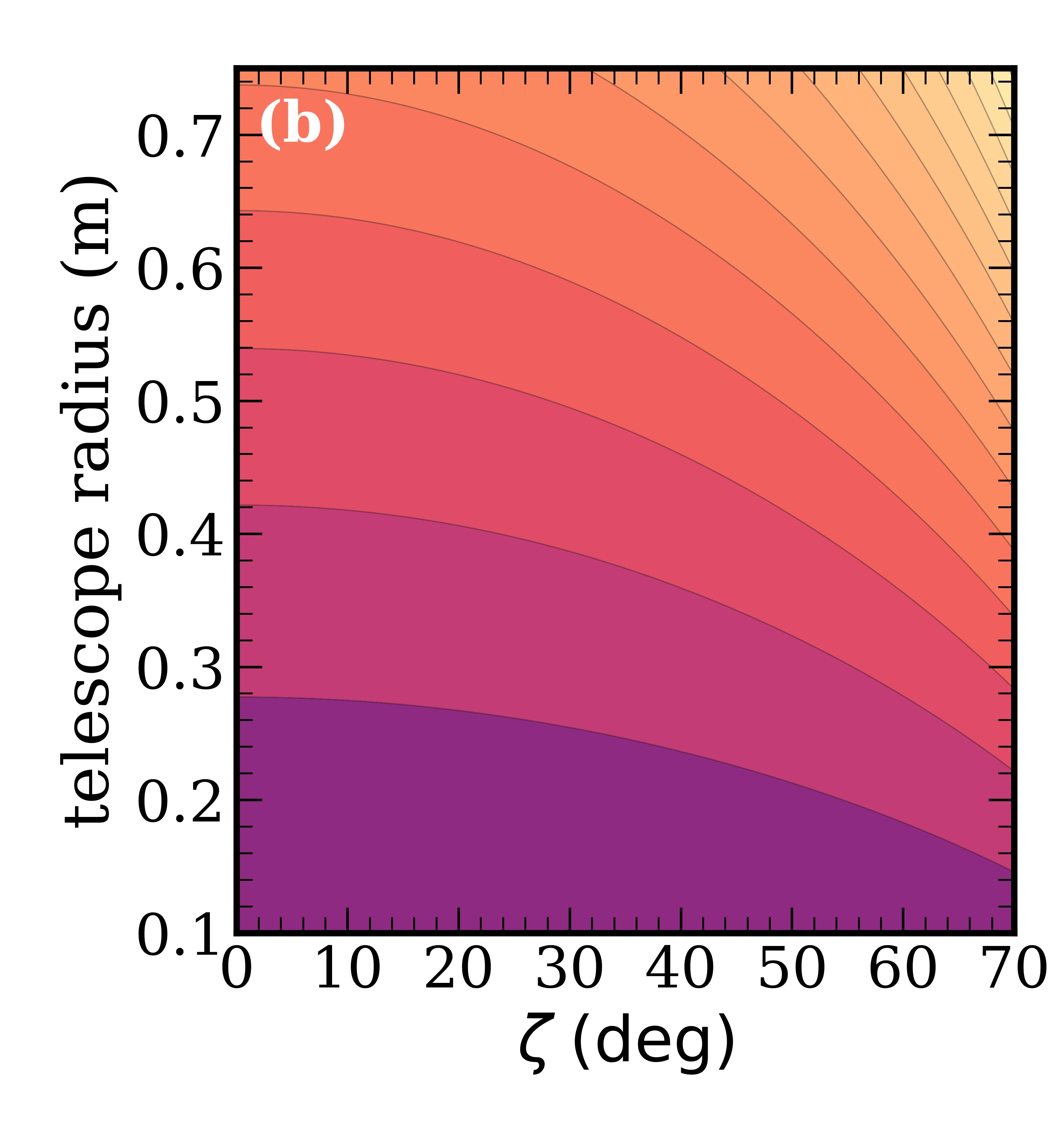}
        \put(4,91){\white\textbf{(b)}}
    \end{overpic}
    \begin{overpic}[height=3.99cm,keepaspectratio]{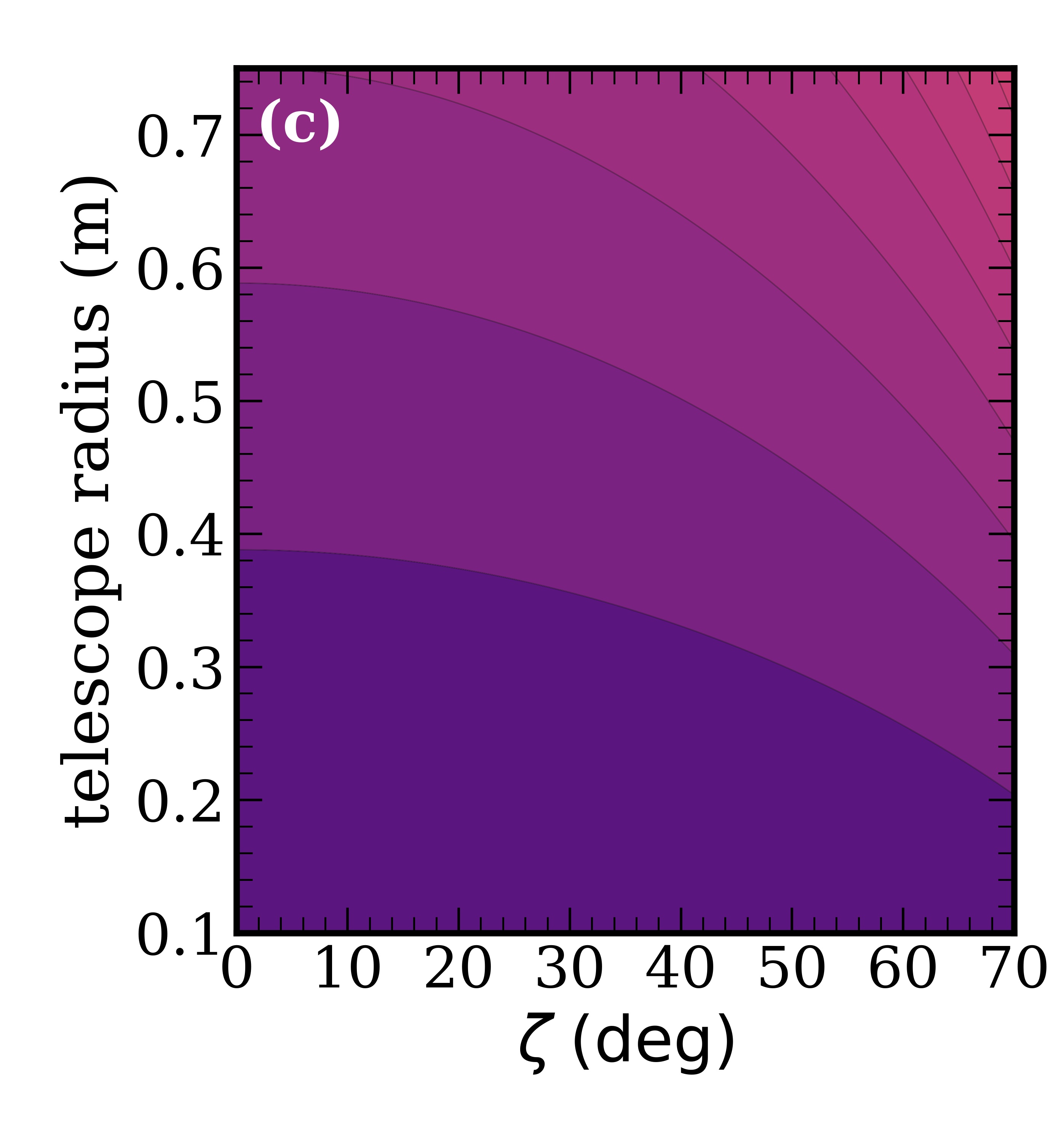}
        \put(4,91){\white\textbf{(c)}}
    \end{overpic}
    \begin{overpic}[height=3.99cm,keepaspectratio]{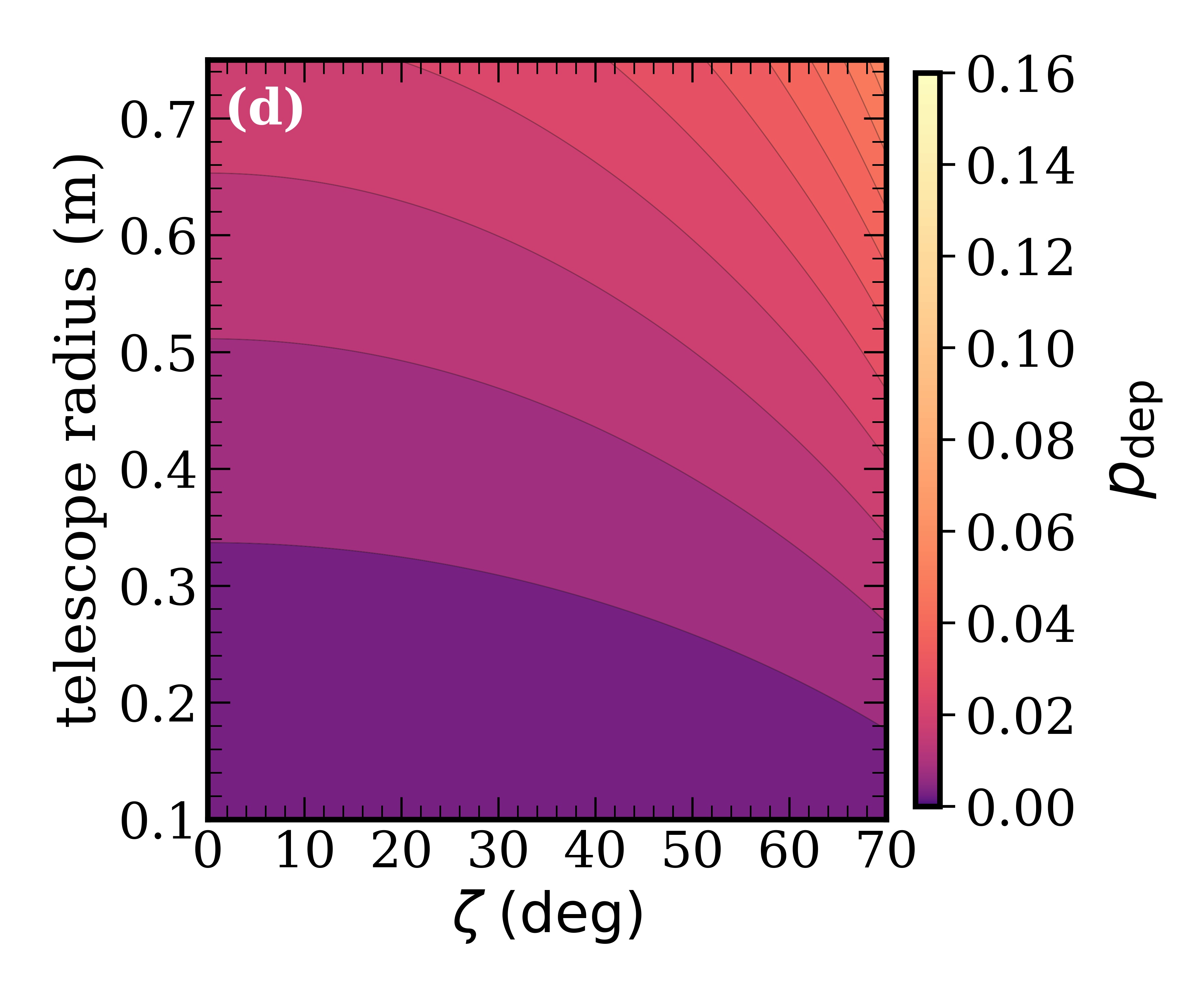}
        \put(4,91){\white\textbf{(d)}}
    \end{overpic}

    \caption{
        Effective depolarization probability $p_{\mathrm{dep}}$ for downlink transmission shown as a function of the telescope radius and the zenith angle. The plot illustrates how larger apertures and larger zenith angles may jointly contribute to effective depolarization probability in the adopted phenomenological model. The results are generated for $\kappa = 4 \times 10^{-3}$, $\lambda=1550$~nm. Panels a–d show the results for different height values (0~m and 2340~m (a.s.l.) corresponding to the Helmos Observatory) and various wind velocity values. The parameters are as follows:
        (a) $H_{\mathrm{OGS}}=0$~m, $v_{\mathrm{rms}}=26.25$~m/s,
        (b) $H_{\mathrm{OGS}}=0$~m, $v_{\mathrm{rms}}=57.00$~m/s,
        (c) $H_{\mathrm{OGS}}=2340$~m, $v_{\mathrm{rms}}=26.25$~m/s,
        (d) $H_{\mathrm{OGS}}=2340$~m, $v_{\mathrm{rms}}=57.00$~m/s.
    }
    \label{fig:DOWN}

\end{figure*}

\begin{figure*}[t]
    \centering

    \centering
    \begin{overpic}[height=3.99cm,keepaspectratio]{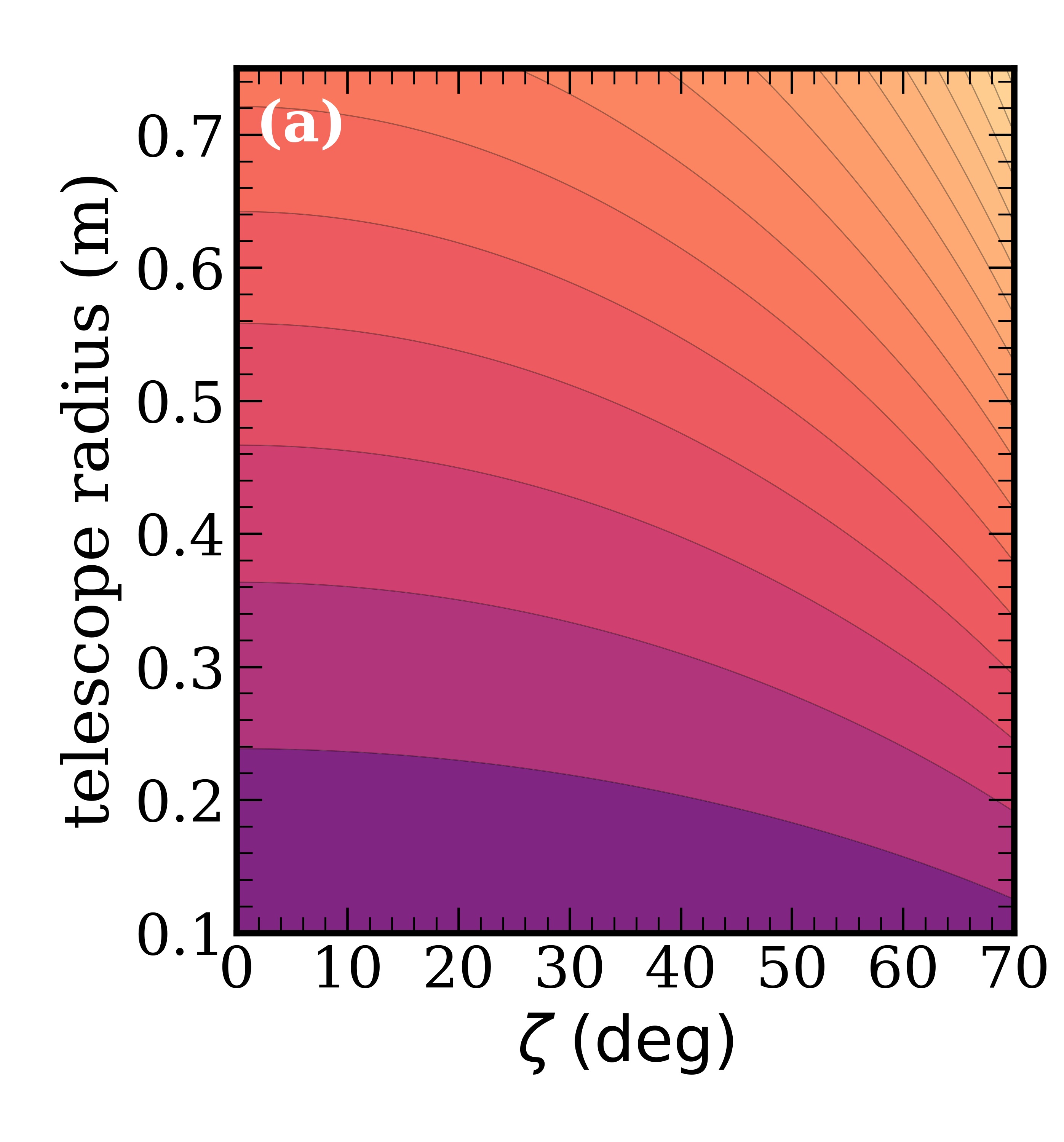}
        \put(4,91){\white\textbf{(a)}}
    \end{overpic}
    \begin{overpic}[height=3.99cm,keepaspectratio]{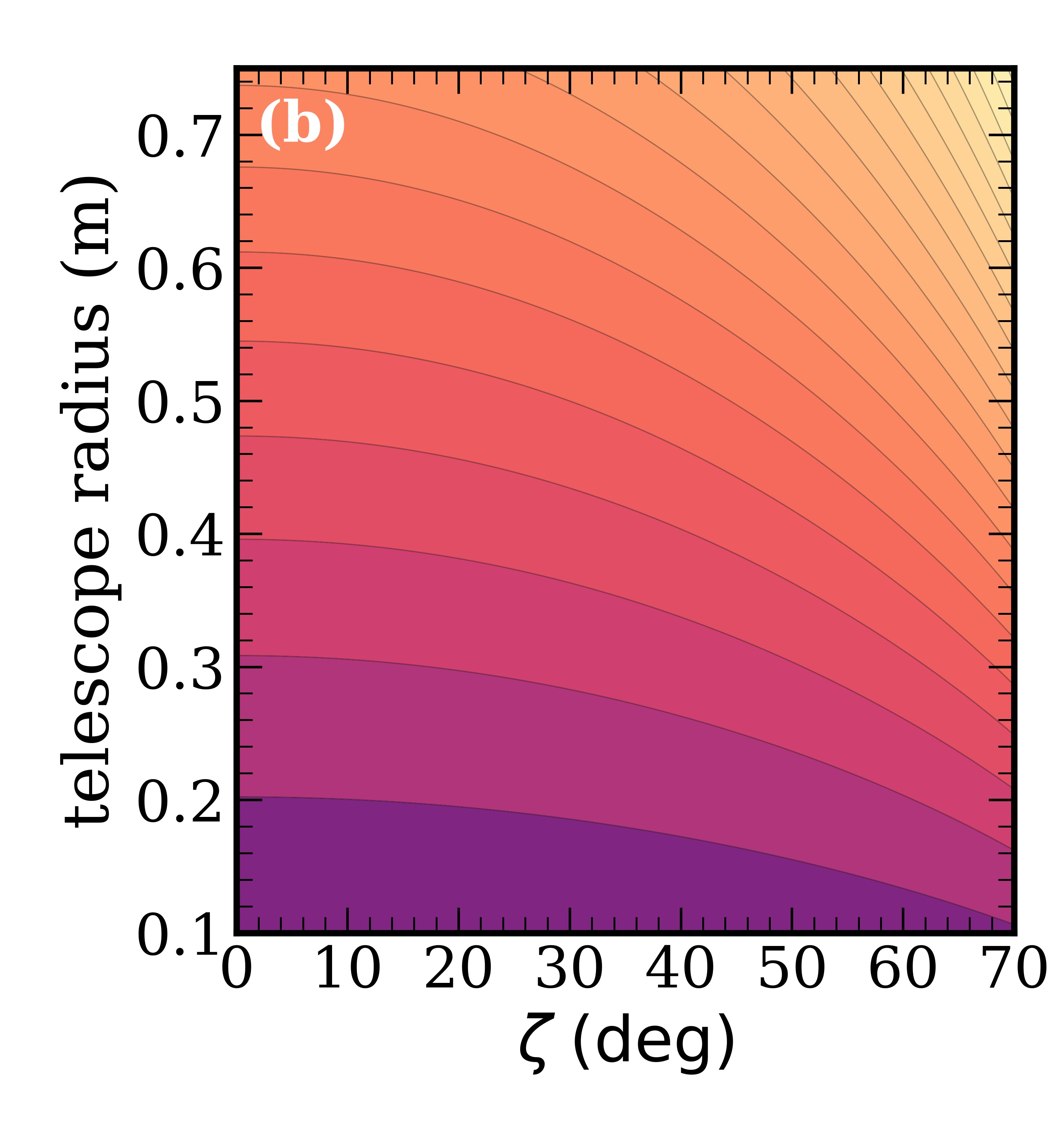}
        \put(4,91){\white\textbf{(b)}}
    \end{overpic}
    \begin{overpic}[height=3.99cm,keepaspectratio]{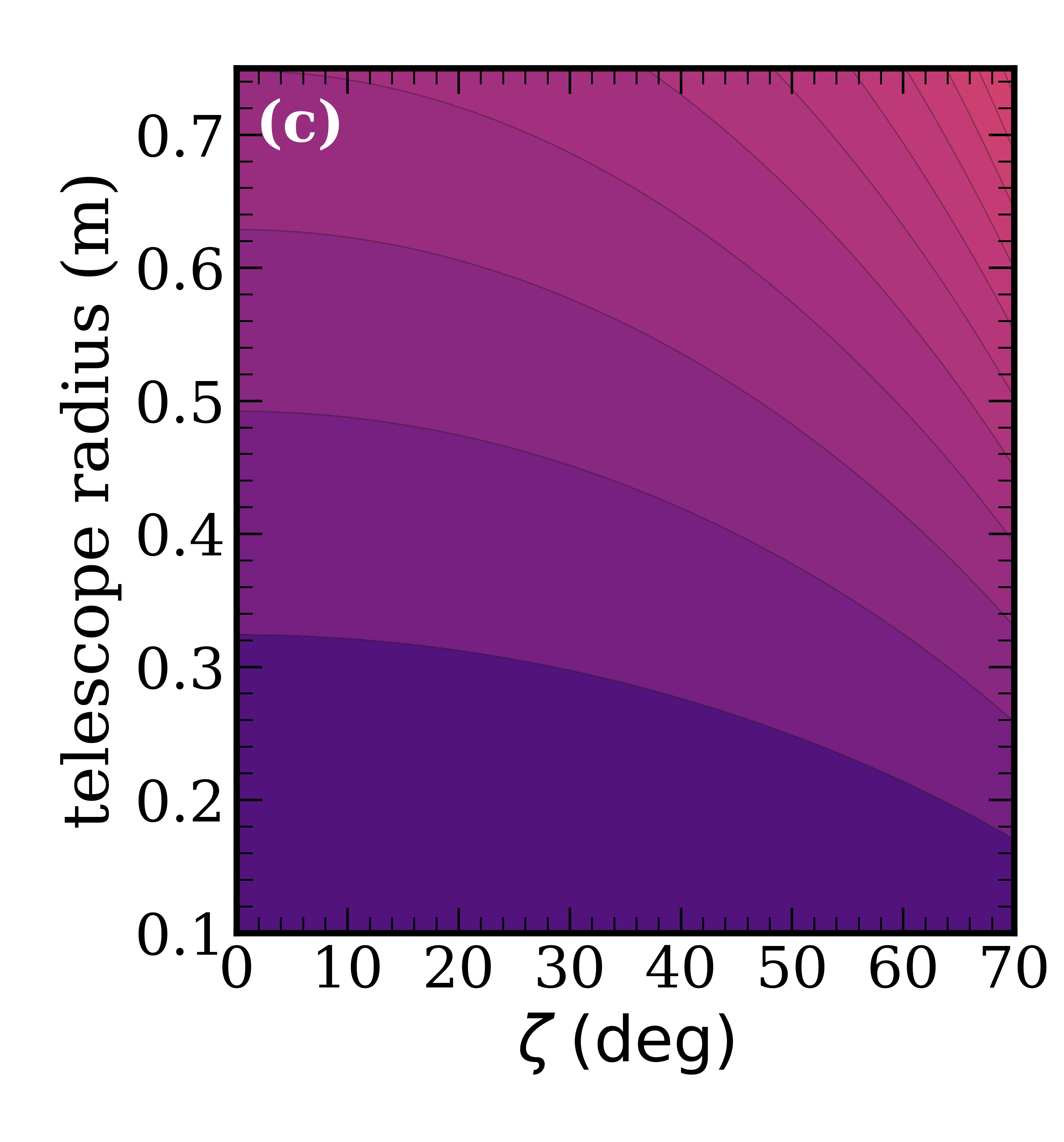}
        \put(4,91){\white\textbf{(c)}}
    \end{overpic}
    \begin{overpic}[height=3.99cm,keepaspectratio]{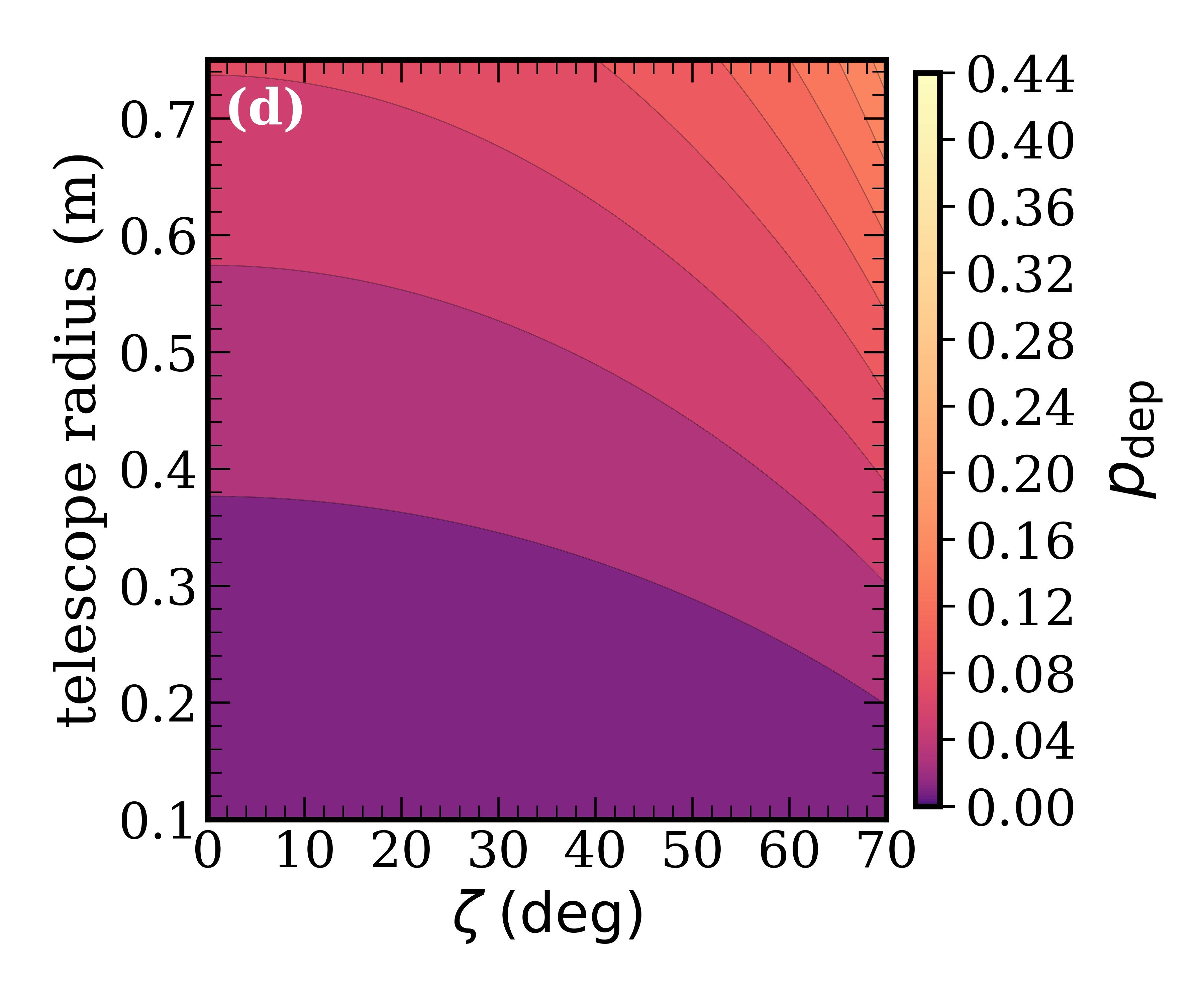}
        \put(4,91){\white\textbf{(d)}}
    \end{overpic}

    \caption{
        Effective depolarization probability $p_{\mathrm{dep}}$ for uplink transmission shown as a function of the telescope radius and the zenith angle. The plot illustrates how larger apertures and larger zenith angles may jointly contribute to effective depolarization probability in the adopted phenomenological model. The results are generated for $\kappa = 4 \times 10^{-3}$, $\lambda=1550$~nm. Panels a–d show the results for different height values (0~m and 2340~m (a.s.l.) corresponding to the Helmos Observatory) and various wind velocity values. The parameters are as follows:
        (a) $H_{\mathrm{OGS}}=0$~m, $v_{\mathrm{rms}}=26.25$~m/s,
        (b) $H_{\mathrm{OGS}}=0$~m, $v_{\mathrm{rms}}=57.00$~m/s,
        (c) $H_{\mathrm{OGS}}=2340$~m, $v_{\mathrm{rms}}=26.25$~m/s,
        (d) $H_{\mathrm{OGS}}=2340$~m, $v_{\mathrm{rms}}=57.00$~m/s.
    }
    \label{fig:UP}

\end{figure*}

Figures~\ref{fig:DOWN} and \ref{fig:UP} illustrate the effective
depolarization probability as a function of telescope radius and
zenith angle for downlink and uplink transmission, respectively,
with $\epsilon=5/3$, $\kappa=4\times10^{-3}$, and
$\lambda=1550$~nm. Both figures exhibit a monotonic increase in
$p_{\mathrm{dep}}$ with increasing aperture radius and zenith
angle. Within the adopted aperture-limited approximation, increasing
the telescope radius increases the effective transverse scale
relative to the atmospheric coherence length. When spatial
distortions are coupled to polarization through the optical system,
this leads to stronger effective polarization mixing. Increasing
the zenith angle further enhances this effect by increasing the
slant path through the turbulent atmosphere and reducing the
Fried parameter.

For a sea-level station and $v_{\mathrm{rms}}=26.25$~m/s, Fig.~\ref{fig:DOWN}(a) predicts depolarization probabilities approaching $0.1$ near the largest aperture radii and zenith angles shown. Increasing the wind
parameter to $57$~m/s raises the upper values to approximately $0.15$ in Fig.~\ref{fig:DOWN}(b). This behavior follows from the wind-dependent contribution to the adopted H--V atmospheric profile, which increases the integrated turbulence strength and therefore decreases $r_0$. The wind parameter should be understood
as a parameter of the atmospheric turbulence profile, rather than as a direct measure of polarization scrambling or local wind
at the telescope.

The station altitude substantially affects the predicted degree of depolarization. To illustrate this dependence, we compare an OGS at sea level with one at \(2340\)~m above sea level, corresponding to the altitude of the Helmos Observatory in Greece. These results are predictions based on the adopted atmospheric profiles and models and do not constitute a site-specific experimental characterization. Nevertheless, raising the OGS from sea level to $2340$~m reduces the largest downlink values from approximately $0.1$ to $0.02$
for $v_{\mathrm{rms}}=26.25$~m/s, as seen by comparing
Figs.~\ref{fig:DOWN}(a) and \ref{fig:DOWN}(c). This corresponds to an approximate $80\%$ reduction in the maximum predicted depolarization. For the higher wind
parameter, the corresponding upper values decrease from
approximately $0.15$ to $0.055$
[Figs.~\ref{fig:DOWN}(b) and \ref{fig:DOWN}(d)]. This corresponds to a reduction of approximately $63\%$. Thus, although the absolute decrease is slightly larger for the higher value of the wind parameter $v_{\mathrm{rms}}$ ($0.095$ compared with $0.08$), the relative reduction in the maximum predicted depolarization is smaller than in the lower-$v_{\mathrm{rms}}$ case.
Within the adopted altitude-dependent profile, this improvement reflects both the exclusion of atmospheric layers below the station and the reduced modeled strength of the local near-ground contribution. The resulting increase in coherence length suppresses effective polarization mixing over the entire range of aperture radii and zenith angles considered.

The relative sensitivity to the wind parameter is more evident at the elevated station. Comparing panels (c) and (d), the upper downlink probabilities increase by approximately a factor of three, whereas the relative increase between the sea-level panels (a) and (b) is smaller. This behavior is consistent with a greater relative contribution of the wind-dependent high-altitude turbulence term after the near-ground contribution has been reduced. Elevating the station therefore improves the predicted polarization quality, but does not eliminate sensitivity to turbulence in the remaining atmospheric path.

The uplink calculations show the same qualitative dependence on
aperture radius, zenith angle, station altitude, and wind parameter.
However, the adopted uplink coherence model produces larger
effective depolarization probabilities than the downlink model
under otherwise identical parameter choices. At sea level, the
largest values are approximately $0.34$ and $0.42$ for the lower
and higher wind parameters, respectively
[Figs.~\ref{fig:UP}(a) and \ref{fig:UP}(b)].
At $2340$~m, these values decrease to approximately $0.06$ and
$0.18$ [Figs.~\ref{fig:UP}(c) and \ref{fig:UP}(d)].
These changes correspond to approximate reductions of $82\%$ and $57\%$ in the maximum predicted depolarization for the lower and higher values of $v_{\mathrm{rms}}$, respectively. In this case, both the relative reduction and the absolute decrease are larger for the lower-$v_{\mathrm{rms}}$ case: the absolute decrease is approximately $0.28$, compared with $0.24$ for the higher wind parameter. Thus, the benefit of an elevated ground station persists in both transmission geometries, while the uplink retains a larger
effective noise contribution in the present calculations.

This directional comparison is conditional on the adopted definitions of the uplink and downlink coherence lengths. Because the same $\kappa$ and effective receiver scale are used in both cases, the numerical difference between the two figures originates from the respective values of $r_0$. An actual comparison of ground-based and satellite-based receivers may also require different coupling coefficients and mode-weighting functions. The plotted values should therefore be interpreted
as a comparison within a common effective-channel
parameterization.

Finally, the aperture dependence describes polarization quality within the assumed receiver model and does not by itself determine the optimal telescope size. A larger aperture can increase the
collected signal while also increasing the effective polarization mixing predicted here. Evaluating this trade-off requires combining the present results with the link budget, coupling efficiency, background counts, and the performance metric of the quantum
protocol. Likewise, the absolute probabilities depend on the chosen value of $\kappa$, which requires calibration for a particular optical system. The figures consequently illustrate the parameter sensitivity of the effective model rather than
universal atmospheric depolarization levels.

\begin{figure}[t]
    \centering
    \includegraphics[width=0.99\columnwidth]{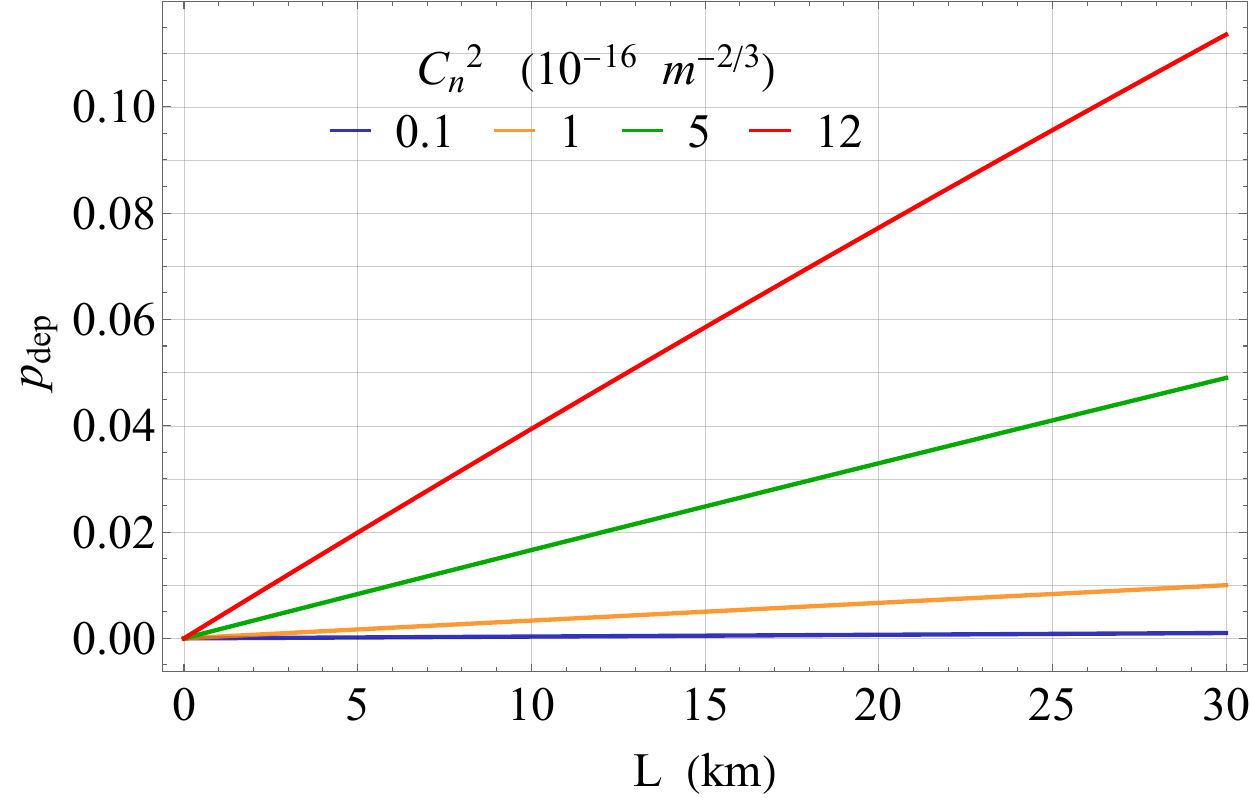}\put(-194,133){(a)}
    \vspace{0.5em}
    \includegraphics[width=0.99\columnwidth]{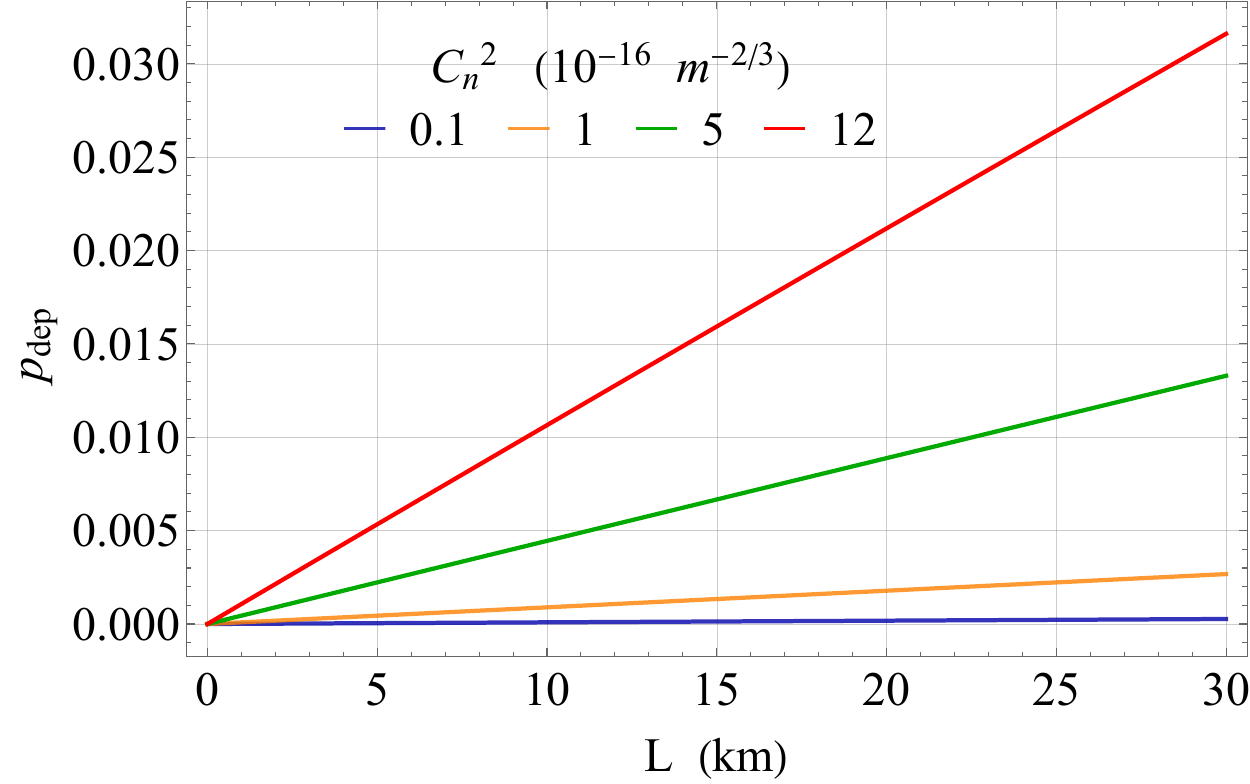}\put(-190,131){(b)}
    \caption{Effective depolarization probability as a function of propagation distance for horizontal free-space optical links, evaluated for $\kappa = 4\times10^{-3}$ and a telescope diameter of $D=25.4$ cm, for two different wavelengths: (a) $\lambda = 800$ nm and (b) $\lambda=1550$ nm.}
    \label{figure04b}
\end{figure}

\begin{figure}[t]
    \centering
    \includegraphics[width=0.99\columnwidth]{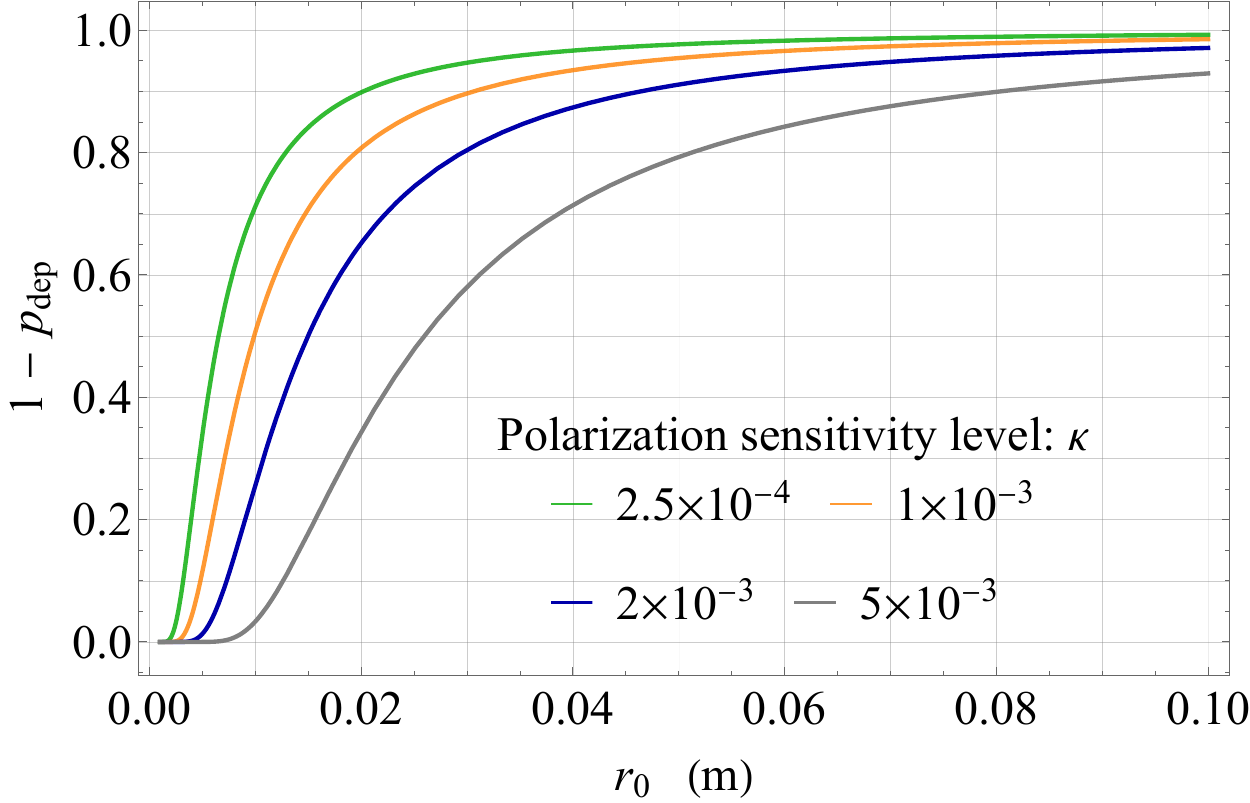}
    \caption{Polarization preservation probability for downlink transmission as a function of the Fried parameter for four polarization sensitivity levels, with the telescope radius fixed at $0.5$\,m.}
    \label{figure05}
\end{figure}

To compare downlink transmission with near-horizontal free-space propagation, we evaluate Eq.~\eqref{eq:pdep_general} using the Fried parameter from Eq.~\eqref{eq:r0_const} under the assumption of constant $C_n^2$. Four representative turbulence strengths are considered:
\[
C_n^2 = 10^{-17},\; 10^{-16},\; 5\times10^{-16},\; 1.2\times10^{-15}\ \mathrm{m}^{-2/3}.
\]
The polarization sensitivity parameter is fixed at $\kappa = 4\times 10^{-3}$, matching the value used in Figs.~\ref{fig:DOWN} and \ref{fig:UP}. For the receiver, we take a telescope diameter of $D = 25.4$ cm, corresponding to a commercially available aperture suitable for short-range FSO tests. The resulting effective depolarization probability is shown in Fig.~\ref{figure04b} as a function of propagation distance $L$, for two different optical wavelengths: (a) $\lambda = 800$ nm and (b) $\lambda=1550$ nm.

Since $r_0$ decreases with increasing $C_n^2L$, the probability $p_{\mathrm{dep}}$ increases with the propagation distance and turbulence strength. For weak turbulence, associated with calm nighttime conditions and values of $C_n^2$ in the range $10^{-17}$--$10^{-16}\ \mathrm{m}^{-2/3}$, the depolarization probability remains below $0.01$ up to $L \approx 30$ km for $\lambda = 800$. In contrast, for stronger turbulence with $C_n^2 = 1.2\times10^{-15}\ \mathrm{m}^{-2/3}$, the model predicts values approaching $p_{\mathrm{dep}}\approx 0.1$ at $L\approx25$ km, which can significantly affect polarization-based quantum communication protocols.

Comparing panels (a) and (b) of Fig.~\ref{figure04b} shows that, for the same link length and turbulence strength, increasing the wavelength from $800$~nm to $1550$~nm substantially reduces the effective depolarization probability. This wavelength dependence follows from the Fried parameter, which scales as $r_0\propto\lambda^{6/5}$. For a fixed receiver aperture, the longer wavelength therefore yields a smaller ratio $a_{\mathrm{eff}}/r_0$ and, consequently, weaker polarization mixing in the model.

Figure~\ref{figure05} shows the polarization retention factor
\begin{equation}
P_{\mathrm{pres}}
\equiv 1-p_{\mathrm{dep}}
=
\exp\!\left[
-\kappa\left(\frac{a_{\mathrm{eff}}}{r_0}\right)^{5/3}
\right]
\end{equation}
as a function of the downlink Fried parameter $r_0$ for four
receiver sensitivity levels, with $a_{\mathrm{eff}}=0.5$~m.
For a detected photon with nonzero input polarization,
$P_{\mathrm{pres}}$ gives the fraction of the initial
Bloch-vector length retained at the receiver. At fixed $r_0$,
a larger $\kappa$ gives a smaller retention factor because the
model assigns a greater polarization-rotation variance to the
same spatial phase disturbance. At fixed $\kappa$, increasing
$r_0$ reduces the phase variation across the aperture, and
$P_{\mathrm{pres}}$ approaches unity as $r_0$ increases.

A qualitatively similar trend appears in Fig.~4 of
Ref.~\cite{Wang2011}, where the quantum degree of polarization
increases with the atmospheric coherence length $\rho_0$ and
approaches a limiting value. The comparison concerns the
dependence on spatial coherence. Wang et al. evaluate a
Stokes-based polarization observable for a specified optical
field, whereas Fig.~\ref{figure05} evaluates the contraction
of a single-photon polarization qubit for different receiver
sensitivities.

\subsection{Practical Notes on Parameters $\kappa$ and $\epsilon$}

The exponent $\epsilon$ is typically chosen close to $5/3$ when the effective depolarization is assumed to follow Kolmogorov-type phase statistics. In this case, the term $(a_{\mathrm{eff}}/r_0)^{5/3}$ is connected to the Kolmogorov phase structure function evaluated at the characteristic spatial scale of the receiver \cite{Kolmogorov1941,Kolmogorov1941b,Tatarski1961,Kolmogorov1991,Benzi2023}. This scaling reflects the mean-square phase difference across a transverse separation and is central to the standard theory of atmospheric wavefront distortions. It is also widely used in astronomical seeing and free-space optical communication, where the ratio between the receiver aperture and the Fried parameter determines the spatial coherence properties of the collected field \cite{Dayton1992,seeing2025}.

The prefactor $\kappa$ quantifies the sensitivity of the complete optical system to spatial perturbations. It describes how efficiently turbulence-induced spatial distortions are converted into effective polarization mixing. Its value depends on the optical configuration, including telescope geometry, mirror coatings, off-axis propagation, polarization-maintaining elements, fiber coupling, detector geometry, and alignment stability. The coefficient $\kappa$ may also absorb residual environmental effects not explicitly captured by the Fried parameter, such as scattering by aerosols, humidity-related variations, or thermal gradients near the receiver.

In practice, $\kappa$ should be determined by calibration. This can be done experimentally by measuring polarization visibility or process fidelity for known atmospheric conditions, or numerically by using phase-screen propagation simulations combined with a model of the receiver optics. For polarization-robust systems, $\kappa$ can be very small, for example on the order of $10^{-3}$ to $10^{-2}$. Less optimized systems may exhibit larger values. Therefore, in the present framework, $\epsilon$ determines the turbulence-scaling law, while $\kappa$ remains the key scenario-dependent parameter.

Several theoretical and experimental works indicate that atmospheric turbulence can influence polarization quality in free-space optical channels through spatial coherence effects, receiver averaging, or coupling between spatial and polarization degrees of freedom \cite{Hohn1969,Bradford1969,Wang2013,Zhang2014,Nikulin2024}. In particular, theoretical studies of the degree of polarization of quantum light in turbulent media show that the Fried parameter can appear as the coherence-length variable controlling the strength of polarization degradation \cite{Wang2011}. Our model can be regarded as a simplified reduction of such more general coherence-integral descriptions: instead of evaluating the full spatial coherence integral, we use a compact exponential law expressed through the dimensionless ratio $a_{\mathrm{eff}}/r_0$. This simplification is suitable for numerical studies of distributed quantum protocols, where the atmospheric link must be represented by a small number of effective channel parameters.

\section{Distributed Quantum Computing: Remote CNOT Gate Protocol Using LOCC and Shared Entanglement}\label{remotecnotgate}

\subsection{Two-Qubit Depolarizing Channel}\label{dualchannel}

In this part of the model, we focus on the effective depolarizing contribution to the atmospheric quantum channel. In satellite-based distribution of polarization-entangled photon pairs, several physical effects may limit the quality of the distributed resource, including propagation loss, beam wandering, pointing errors, intensity fluctuations, background counts, and detector imperfections. Here, we isolate the effect of polarization noise and describe it by an effective two-qubit depolarizing channel. Since each photon propagates through an independent optical path from the satellite to its corresponding OGS, we assume that each qubit undergoes depolarization independently. Let $\rho_{ab}$ be an arbitrary two-qubit density matrix describing the polarization state of the photon pair, and denote the depolarization probabilities in the first and second optical paths by $p$ and $q$, respectively.

For a single qubit, the depolarizing channel can be written as \cite{Nielsen2000}
\begin{equation}
\mathcal{E}_p(\rho_1) = (1-p)\,\rho_1 + \frac{p}{2}\,\mathbb{I}_2,
\end{equation}
where $\mathbb{I}_2$ is the $2\times 2$ identity matrix and $\rho_1$ is the single-qubit density matrix. This channel maps the input state to itself with probability $1-p$ and replaces it by the maximally mixed state with probability $p$.

For two qubits, the independent action of two depolarizing channels is described by the product channel
\[
\mathcal{E}_{p,q} = \mathcal{E}_{p} \otimes \mathcal{E}_{q}.
\]
The corresponding output density matrix can be expressed as
\begin{equation}
\label{2qubitdepol}
\begin{aligned}
\mathcal{E}_{p,q}(\rho_{ab}) 
=&\; (1-p)(1-q)\,\rho_{ab}\\  
&+ (1-p)\frac{q}{2}\,
\big(\mathrm{Tr}_{b}[\rho_{ab}] \otimes \mathbb{I}_2 \big)\\ 
&+ \frac{p}{2}(1-q)\,
\big(\mathbb{I}_2 \otimes \mathrm{Tr}_{a}[\rho_{ab}]\big)\\ 
&+ \frac{pq}{4}\,\mathbb{I}_4 ,
\end{aligned}
\end{equation}
where $\mathrm{Tr}_{a}$ and $\mathrm{Tr}_{b}$ denote the partial trace over the first and second qubits, respectively, and $\mathbb{I}_4$ is the $4\times 4$ identity matrix. The second term describes the case in which only the second qubit is depolarized, while the first qubit is preserved. The third term describes the opposite case, where only the first qubit is depolarized. The last term corresponds to simultaneous depolarization of both qubits and gives the maximally mixed two-qubit contribution.

This formulation preserves the trace and positivity of the density matrix and provides a direct extension of the standard single-qubit depolarizing channel to two spatially separated photons affected by independent atmospheric channels.

For numerical simulations, the same map can be implemented using Kraus operators. Denoting the Pauli matrices by $\sigma_x$, $\sigma_y$, and $\sigma_z$, the Kraus representation of the single-qubit depolarizing channel acting on the first qubit is \cite{Nielsen2000}
\begin{equation}
K_0 = \sqrt{1 - \frac{3}{4}p}\, \mathbb{I}_2,
\qquad
K_i = \sqrt{\frac{p}{4}}\, \sigma_i,
\end{equation}
where $i=x,y,z$. Analogously, the channel acting on the second qubit is described by operators $L_0$, $L_x$, $L_y$, and $L_z$, obtained by replacing $p$ with $q$. The full two-qubit channel is then represented by the set of $16$ Kraus operators
\begin{equation}
M_{ij} = K_i \otimes L_j,
\qquad i,j \in \{0,x,y,z\},
\end{equation}
so that
\begin{equation}
\mathcal{E}_{p,q}(\rho_{ab})
=
\sum_{i,j}
M_{ij}\rho_{ab}M_{ij}^{\dagger}.
\end{equation}

\subsection{Theoretical Framework}\label{remotecnotgateth}

We consider a protocol for implementing a remote controlled-NOT (CNOT) gate between two spatially separated quantum processors using only local operations and classical communication (LOCC), assisted by a pre-shared Bell pair. Such a protocol constitutes one of the fundamental building blocks (primitives) of DQC, as it enables the realization of entangling gates between qubits that do not physically interact \cite{Eisert2000,Duan2004,Jiang2007} Together with arbitrary local single-qubit gates, remote CNOT gates provide a universal gate set for quantum computation across distributed quantum registers.

\begin{figure}[h!]
    \centering
    \includegraphics[width=0.99\columnwidth]{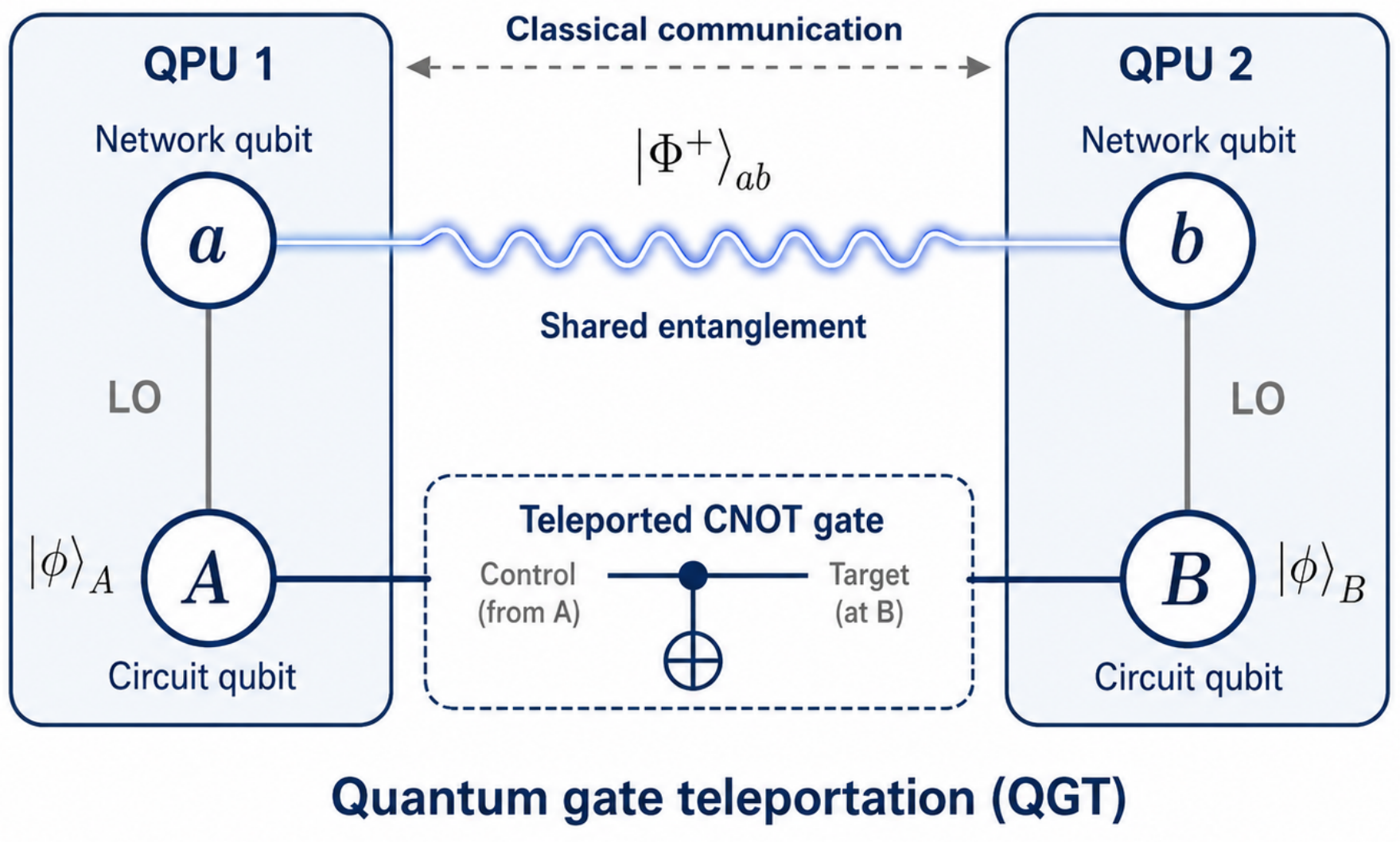}
    \caption{Conceptual illustration of quantum gate teleportation with four qubits located in two quantum processing units, cf. Ref.~\cite{main2025distributed}.}
    \label{figure00g2}
\end{figure}

Figure~\ref{figure00g2} presents the minimal four-qubit architecture required for a QGT protocol based in Ref~\cite{main2025distributed}. The two spatially separated QPUs, introduced in Fig.~\ref{figure00g1}, each consist of one network qubit and one circuit qubit. The network qubits, denoted by \(a\) and \(b\), are initialized in the maximally entangled Bell state
\begin{equation}\label{bellstateeq}
    \ket{\Phi^+_{ab}}=
    \frac{1}{\sqrt{2}}
    \left(
    \ket{0}_a\ket{0}_b+
    \ket{1}_a\ket{1}_b
    \right),
\end{equation}
which serves as the nonlocal quantum resource required for the protocol. The circuit qubits, labeled \(A\) and \(B\), carry the input states \(\ket{\phi}_A\) and \(\ket{\phi}_B\), respectively, and represent the logical qubits on which the distributed CNOT operation is performed.

Within each QPU, the network qubit is coupled locally to the corresponding circuit qubit through local operations (LO), while a classical communication channel is used to transmit measurement outcomes and feed-forward correction information between the two processors. In this setting, QGT implements a nonlocal CNOT gate, with circuit qubit \(A\) acting as the control qubit and circuit qubit \(B\) as the target qubit. Consequently, the combination of the shared entangled pair \(\ket{\Phi^{+}}_{ab}\) and LOCC enables an effective two-qubit gate between remote quantum processors, establishing QGT as a fundamental primitive for DQC.

\subsubsection*{Initial Setup}

We consider a four-qubit setting consisting of two circuit qubits, \(A\) and \(B\), and two ancillary network qubits, \(a\) and \(b\). Qubit \(A\), located in module~1, plays the role of the control qubit and is initially prepared in the state
\begin{equation}
    \ket{\phi}_A = \alpha \ket{0}_A + \beta \ket{1}_A,
\end{equation}
whereas qubit \(B\), located in module~2, plays the role of the target qubit and is initially prepared in the state
\begin{equation}
    \ket{\phi}_B = \gamma \ket{0}_B + \delta \ket{1}_B.
\end{equation}
The complex amplitudes satisfy the normalization conditions
\begin{equation}
    |\alpha|^2+|\beta|^2=1,
    \qquad
    |\gamma|^2+|\delta|^2=1.
\end{equation}
The two modules share an entangled Bell pair of ancillary network qubits \(a\) and \(b\), defined as in Eq.~(\ref{bellstateeq}).

In practice, the entangled resource may be affected by noise. A realistic description can account for this effect by applying a two-qubit depolarizing channel to the shared Bell state, as defined in Eq.~(\ref{2qubitdepol}). The parameters \(p\) and \(q\) denote the depolarization probabilities on Alice's and Bob's sides, respectively, and can be modeled for an FSO link using Eq.~(\ref{eq:pdep_general}). In the following derivation, however, we first present the ideal protocol in terms of pure states. This provides the reference transformation against which the noisy implementation can later be compared.

The objective of the protocol is to implement the nonlocal gate
\begin{equation}
    \mathrm{CNOT}_{A\rightarrow B},
\end{equation}
where \(A\) is the control qubit and \(B\) is the target qubit. The implementation uses only local operations, local measurements, classical communication, and the pre-shared entangled pair.

\subsubsection*{Global Initial State}

The global initial state, written in the ordering \(AabB\), is given by
\begin{equation}
\begin{split}
    \ket{\Psi_0}
    =&
    \left(
    \alpha \ket{0}_A + \beta \ket{1}_A
    \right)
    \otimes
    \frac{1}{\sqrt{2}}
    \left(
    \ket{0}_a\ket{0}_b+
    \ket{1}_a\ket{1}_b
    \right)\\
    &\otimes
    \left(
    \gamma \ket{0}_B + \delta \ket{1}_B
    \right).
\end{split}
\end{equation}
Equivalently, after expanding the state of qubits \(A\), \(a\), and \(b\), one obtains
\begin{equation}
\begin{split}
    \ket{\Psi_0}
    =
    \frac{1}{\sqrt{2}}
    \Big(
    &\alpha \ket{0}_A\ket{0}_a\ket{0}_b
    +
    \alpha \ket{0}_A\ket{1}_a\ket{1}_b \\ 
    &+
    \beta \ket{1}_A\ket{0}_a\ket{0}_b
    +
    \beta \ket{1}_A\ket{1}_a\ket{1}_b
    \Big)\\
    &\otimes
    \left(
    \gamma \ket{0}_B + \delta \ket{1}_B
    \right).
\end{split}
\end{equation}

\subsubsection*{Step 1: Local CNOT from \(A\) to \(a\)}

Alice first applies a local CNOT gate with circuit qubit \(A\) as the control and network qubit \(a\) as the target. Its action is
\begin{equation}
    \ket{x}_A\ket{y}_a
    \mapsto
    \ket{x}_A\ket{y\oplus x}_a,
    \qquad x,y\in\{0,1\}.
\end{equation}
After this operation, the global state becomes
\begin{equation}
\begin{split}
    \ket{\Psi_1}
    =
    \frac{1}{\sqrt{2}}
    \Big[
    &\ket{0}_A
    \left(
    \alpha \ket{0}_a\ket{0}_b
    +
    \alpha \ket{1}_a\ket{1}_b
    \right)\\  
    &+
    \ket{1}_A
    \left(
    \beta \ket{1}_a\ket{0}_b
    +
    \beta \ket{0}_a\ket{1}_b
    \right)
    \Big]\\
    &\otimes
    \left(
    \gamma \ket{0}_B + \delta \ket{1}_B
    \right).
\end{split}
\end{equation}
This local operation correlates Alice's circuit qubit \(A\) with the network qubit \(a\), thereby transferring the control information into the entangled resource.

\subsubsection*{Step 2: Measurement of Qubit \(a\) in the Computational Basis}

Alice then measures qubit \(a\) in the computational basis \(\{\ket{0},\ket{1}\}\). Let the measurement outcome be denoted by
\begin{equation}
    m_a \in \{0,1\}.
\end{equation}
Conditioned on this outcome, and after dropping the common normalization factor, the post-measurement states are as follows.

For \(m_a=0\), the state is
\begin{equation}
\begin{split}
    \ket{\Psi^{m_a=0}_2}
    =&
    \left(
    \alpha \ket{0}_A\ket{0}_b
    +
    \beta \ket{1}_A\ket{1}_b
    \right)\\
    &\otimes
    \left(
    \gamma \ket{0}_B + \delta \ket{1}_B
    \right).
\end{split}
\end{equation}
For \(m_a=1\), the state is
\begin{equation}
\begin{split}
    \ket{\Psi^{m_a=1}_2}
    =&
    \left(
    \alpha \ket{0}_A\ket{1}_b
    +
    \beta \ket{1}_A\ket{0}_b
    \right)\\
    &\otimes
    \left(
    \gamma \ket{0}_B + \delta \ket{1}_B
    \right).
\end{split}
\end{equation}
Thus, the measurement outcome \(m_a\) determines whether the state of Bob's network qubit \(b\) is correlated or anti-correlated with Alice's control qubit \(A\).

\subsubsection*{Step 3: Local CNOT from \(b\) to \(B\)}

Bob next applies a local CNOT gate with network qubit \(b\) as the control and circuit qubit \(B\) as the target. This operation flips \(B\) whenever \(b=1\).

If \(m_a=0\), the state after Bob's local CNOT operation is
\begin{equation}\label{eq_55}
\begin{split}
    \ket{\Psi^{m_a=0}_3}
    =
    &\alpha \ket{0}_A\ket{0}_b
    \left(
    \gamma \ket{0}_B + \delta \ket{1}_B
    \right)\\ 
    &+
    \beta \ket{1}_A\ket{1}_b
    \left(
    \gamma \ket{1}_B + \delta \ket{0}_B
    \right).
\end{split}
\end{equation}
If \(m_a=1\), the corresponding state is
\begin{equation}\label{eq_56}
\begin{split}
    \ket{\Psi^{m_a=1}_3}
    =
    &\alpha \ket{0}_A\ket{1}_b
    \left(
    \gamma \ket{1}_B + \delta \ket{0}_B
    \right)\\ 
    &+
    \beta \ket{1}_A\ket{0}_b
    \left(
    \gamma \ket{0}_B + \delta \ket{1}_B
    \right).
\end{split}
\end{equation}

\subsubsection*{Step 4: Measurement of Qubit \(b\) in the Hadamard Basis}

Bob then measures qubit \(b\) in the Hadamard basis,
\begin{equation}
    \ket{+}=\frac{1}{\sqrt{2}}\left(\ket{0}+\ket{1}\right),
    \qquad
    \ket{-}=\frac{1}{\sqrt{2}}\left(\ket{0}-\ket{1}\right).
\end{equation}
We denote the measurement result by
\begin{equation}
    m_b\in\{0,1\},
\end{equation}
with the convention
\begin{equation}
    m_b=0 \Longleftrightarrow \ket{+},
    \qquad
    m_b=1 \Longleftrightarrow \ket{-}.
\end{equation}
To express the states before measurement in this basis, we use
\begin{equation}
    \ket{0}_b=
    \frac{1}{\sqrt{2}}
    \left(
    \ket{+}_b+\ket{-}_b
    \right),
    \quad
    \ket{1}_b=
    \frac{1}{\sqrt{2}}
    \left(
    \ket{+}_b-\ket{-}_b
    \right).
\end{equation}

For \(m_a=0\), substituting the Hadamard-basis representation into \(\ket{\Psi^{m_a=0}_3}\) from Eq.~(\ref{eq_55}) gives
\begin{equation}
\begin{split}
    \ket{\Psi^{m_a=0}_3}
    =
   & \frac{1}{\sqrt{2}}\ket{+}_b
    \Big[
    \alpha \ket{0}_A
    \left(
    \gamma \ket{0}_B+\delta \ket{1}_B
    \right)\\ 
    &+
    \beta \ket{1}_A
    \left(
    \gamma \ket{1}_B+\delta \ket{0}_B
    \right)
    \Big] \\&
    +
    \frac{1}{\sqrt{2}}\ket{-}_b
    \Big[
    \alpha \ket{0}_A
    \left(
    \gamma \ket{0}_B+\delta \ket{1}_B
    \right)\\ 
    &-
    \beta \ket{1}_A
    \left(
    \gamma \ket{1}_B+\delta \ket{0}_B
    \right)
    \Big].
\end{split}
\end{equation}
Therefore, if \(m_b=0\), the state of the circuit qubits \(A\) and \(B\) is
\begin{equation}
\begin{split}
    \ket{\psi_{AB}^{00}}
    =
    &\alpha \ket{0}_A
    \left(
    \gamma \ket{0}_B+\delta \ket{1}_B
    \right)\\ 
    &+
    \beta \ket{1}_A
    \left(
    \gamma \ket{1}_B+\delta \ket{0}_B
    \right),
\end{split}
\end{equation}
whereas for \(m_b=1\) it becomes
\begin{equation}
\begin{split}
    \ket{\psi_{AB}^{01}}
    =
    &\alpha \ket{0}_A
    \left(
    \gamma \ket{0}_B+\delta \ket{1}_B
    \right)\\ 
    &-
    \beta \ket{1}_A
    \left(
    \gamma \ket{1}_B+\delta \ket{0}_B
    \right).
\end{split}
\end{equation}

For \(m_a=1\), the state Eq.~(\ref{eq_56}) before the measurement of \(b\) can be rewritten as
\begin{equation}
\begin{split}
    \ket{\Psi^{m_a=1}_3}
    =
    \frac{1}{\sqrt{2}}\ket{+}_b
    \Big[
    &\alpha \ket{0}_A
    \left(
    \gamma \ket{1}_B+\delta \ket{0}_B
    \right)\\ 
    &+
    \beta \ket{1}_A
    \left(
    \gamma \ket{0}_B+\delta \ket{1}_B
    \right)
    \Big] \\
    +
    \frac{1}{\sqrt{2}}\ket{-}_b
    \Big[
    &-\alpha \ket{0}_A
    \left(
    \gamma \ket{1}_B+\delta \ket{0}_B
    \right)\\ 
    &+
    \beta \ket{1}_A
    \left(
    \gamma \ket{0}_B+\delta \ket{1}_B
    \right)
    \Big].
\end{split}
\end{equation}
Thus, for \(m_b=0\), the circuit-qubit state is
\begin{equation}
\begin{split}
    \ket{\psi_{AB}^{10}}
    =
    &\alpha \ket{0}_A
    \left(
    \gamma \ket{1}_B+\delta \ket{0}_B
    \right)\\ 
    &+
    \beta \ket{1}_A
    \left(
    \gamma \ket{0}_B+\delta \ket{1}_B
    \right),
\end{split}
\end{equation}
whereas for \(m_b=1\), it is
\begin{equation}
\begin{split}
    \ket{\psi_{AB}^{11}}
    =
    &-\alpha \ket{0}_A
    \left(
    \gamma \ket{1}_B+\delta \ket{0}_B
    \right)\\ 
    &+
    \beta \ket{1}_A
    \left(
    \gamma \ket{0}_B+\delta \ket{1}_B
    \right).
\end{split}
\end{equation}

\subsubsection*{Step 5: Pauli-Frame Correction}

The desired output state is obtained by applying the CNOT gate from \(A\) to \(B\) to the initial circuit-qubit state:
\begin{equation}
\begin{split}
    \ket{\psi_{\mathrm{CNOT}}}
    &=
    \mathrm{CNOT}_{A\rightarrow B}
    \left(
    \ket{\phi}_A\otimes\ket{\phi}_B
    \right)\\ 
    &=
    \alpha \ket{0}_A
    \left(
    \gamma \ket{0}_B+\delta \ket{1}_B
    \right)\\
    &+
    \beta \ket{1}_A
    \left(
    \gamma \ket{1}_B+\delta \ket{0}_B
    \right).
\end{split}
\end{equation}
Comparing this state with the four conditional states obtained above shows that the measurement outcome \(m_a\) determines whether a bit-flip correction is required on the target qubit \(B\), while the measurement outcome \(m_b\) determines whether a phase correction is required on the control qubit \(A\). Therefore, the required Pauli-frame correction is
\begin{equation}
    Z_A^{m_b} X_B^{m_a},
\end{equation}
up to an irrelevant global phase. Explicitly, the feed-forward correction table is
\begin{equation}
\renewcommand{\arraystretch}{1.4}
\begin{array}{c@{\qquad}c@{\qquad}c}
\hline
m_a & m_b & \text{Correction} \\
\hline
0 & 0 & \mathbb{I}_A\otimes\mathbb{I}_B \\
0 & 1 & Z_A\otimes\mathbb{I}_B \\
1 & 0 & \mathbb{I}_A\otimes X_B \\
1 & 1 & Z_A\otimes X_B \\
\hline
\end{array}
\end{equation}
where the last case recovers the desired state up to a global phase. After the corresponding Pauli-frame update, the final circuit-qubit state is
\begin{equation}\label{cnotideal}
    \ket{\psi_{\mathrm{final}}}
    =
    \mathrm{CNOT}_{A\rightarrow B}
    \left(
    \ket{\phi}_A\otimes\ket{\phi}_B
    \right).
\end{equation}

This derivation shows that a nonlocal entangling gate can be implemented deterministically using local gates, local measurements, classical communication, and one shared Bell pair. In the ideal case, the protocol exactly reproduces the action of \(\mathrm{CNOT}_{A\rightarrow B}\) on the remote circuit qubits. In noisy implementations, the fidelity of the teleported gate is determined primarily by the quality of the distributed Bell pair and by imperfections in the local operations and measurements. This makes the protocol a natural benchmark for distributed quantum computing in modular architectures and quantum networks, including satellite-based systems.

\subsubsection*{Performance Indicators}

To quantify the performance of QGT, we use the standard quantum fidelity between the ideal output state and the state obtained from the simulated protocol. The ideal output is defined as in Eq.~(\ref{cnotideal})
whereas the actual output state produced by the nonlocal implementation of the gate is denoted by \(\rho\). Since the reference state is pure, the fidelity is given by
\begin{equation}\label{fidelitydef}
    F(\psi_{\mathrm{final}},\rho)
    =
    \bra{\psi_{\mathrm{final}}}
    \rho
    \ket{\psi_{\mathrm{final}}}.
\end{equation}
This quantity measures how close the simulated output of the QGT protocol is to the state that would be obtained from an ideal CNOT operation applied directly to the circuit qubits \(A\) and \(B\).

In the ideal case, when the resource state is the maximally entangled Bell state \(\ket{\Phi^{+}}_{ab}\) and no additional noise is present, the protocol exactly reproduces the action of \(\mathrm{CNOT}_{A\rightarrow B}\). Consequently, the fidelity reaches unity, \(F=1\), as follows directly from the step-by-step derivation of the QGT protocol. In a realistic satellite-based implementation, however, the distributed entangled pair is affected by atmospheric propagation and receiver imperfections. In our model, these effects are represented by a two-qubit depolarizing channel, as introduced in Eq.~(\ref{2qubitdepol}), while the corresponding depolarization probabilities, $p$ and $q$, are determined by atmospheric and optical parameters according to Eq.~(\ref{eq:pdep_general}).

The resulting fidelity therefore becomes a function of the depolarization probabilities \(p\) and \(q\), which describe the degradation of the entangled resource on Alice's and Bob's sides, respectively. The corresponding numerical correspondence \((p,q) \rightarrow F(\psi_{\mathrm{final}},\rho)\) can be derived symbolically, for example with the assistance of Wolfram Mathematica, and is then used as the figure of merit for assessing the performance of QGT under different channel conditions.

\subsection{QGT Fidelity Under Different Atmospheric Conditions and Receiver Apertures}\label{remotecnotgateres}

\begin{figure*}[t]
    \centering
    \begin{minipage}[t]{0.49\linewidth}
        \centering
        \includegraphics[height=6.8cm, keepaspectratio]{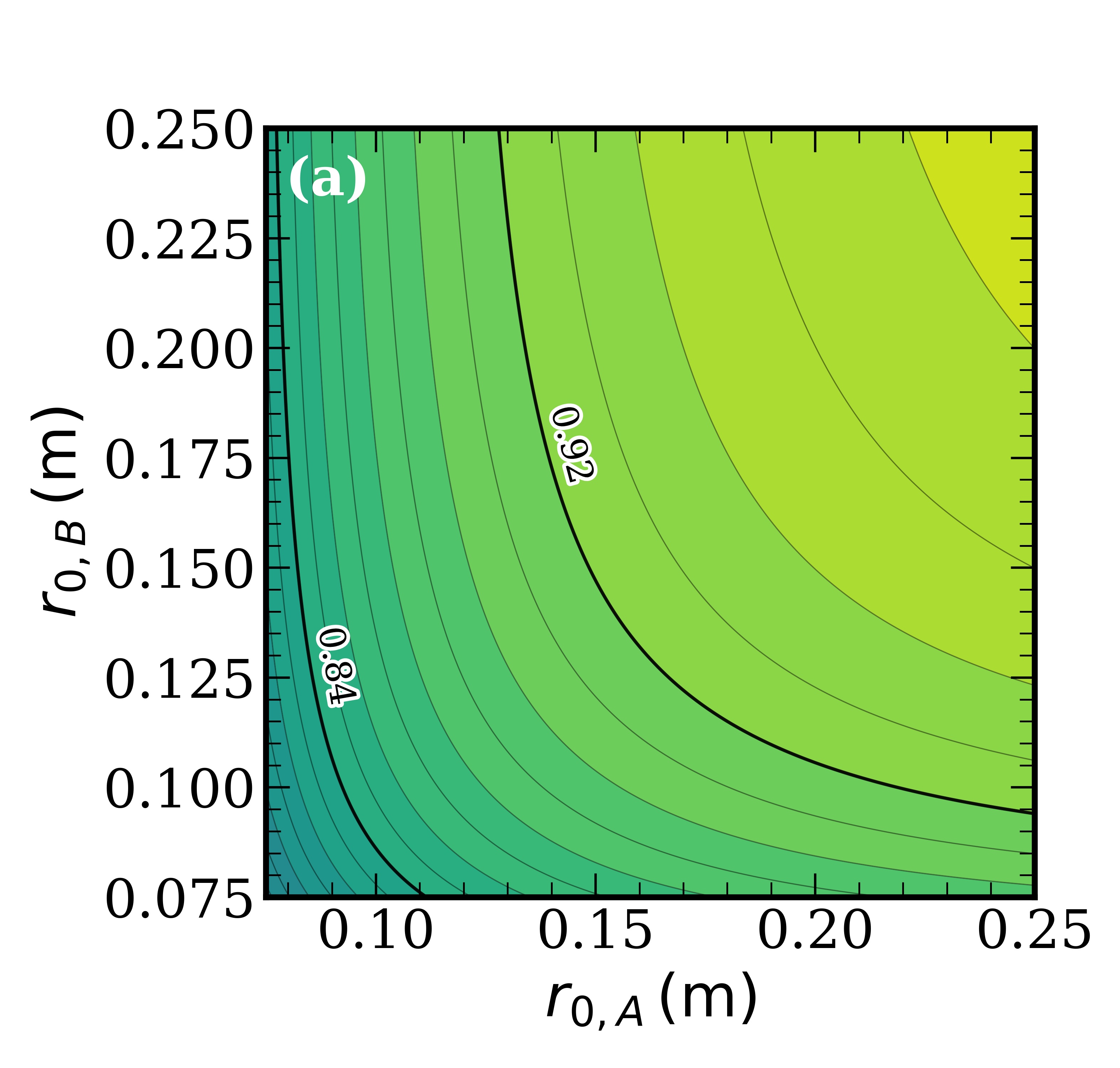}
    \end{minipage}\hfill
    \begin{minipage}[t]{0.49\linewidth}
        \centering
        \includegraphics[height=6.8cm, keepaspectratio]{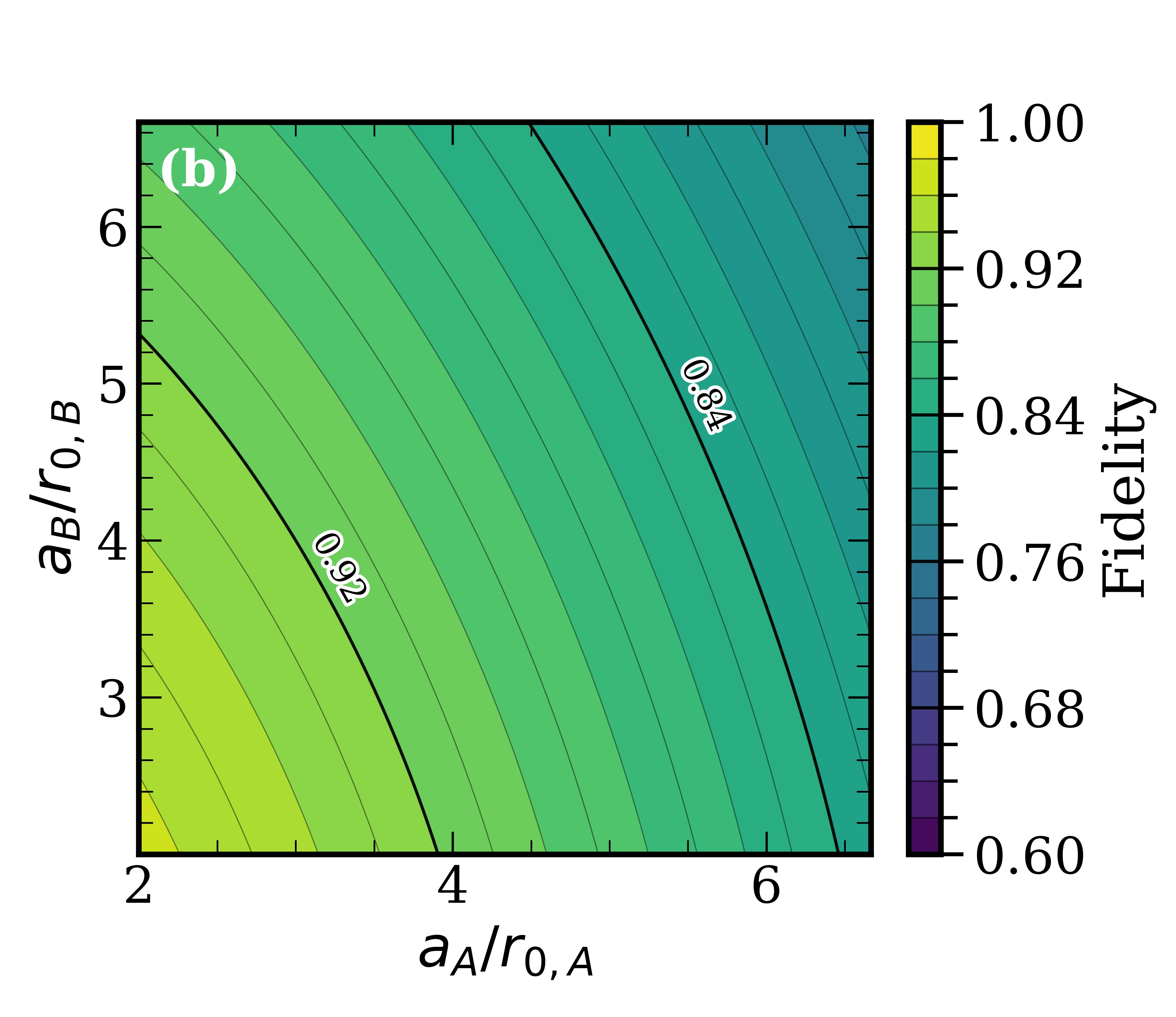}
    \end{minipage}
    \caption{Scenario I: asymmetric turbulence parameters $\kappa_A=0.01$ and $\kappa_B=0.005$. Panel (a) shows the dependence of QGT fidelity on the Fried parameters $r_{0,A}$ and $r_{0,B}$ for equal apertures: $a_A=a_B=0.5\,\mathrm{m}$. Panel (b) presents the same dependence in terms of the dimensionless ratios $a_A/r_{0,A}$ and $a_B/r_{0,B}$. The states coefficients are: $\alpha=\frac{1}{\sqrt{2}}$, $ \beta = \sqrt{1 - {\lvert \alpha  \rvert}^2}$, $\gamma = \frac{i}{\sqrt{2}}$, $\delta = \sqrt{1- {\lvert \gamma \rvert}^2}$.}
    \label{fig:SCENARIO_A}
\end{figure*}

\begin{figure*}[t]
    \centering
    \begin{minipage}[t]{0.49\linewidth}
        \centering
        \includegraphics[height=6.8cm, keepaspectratio]{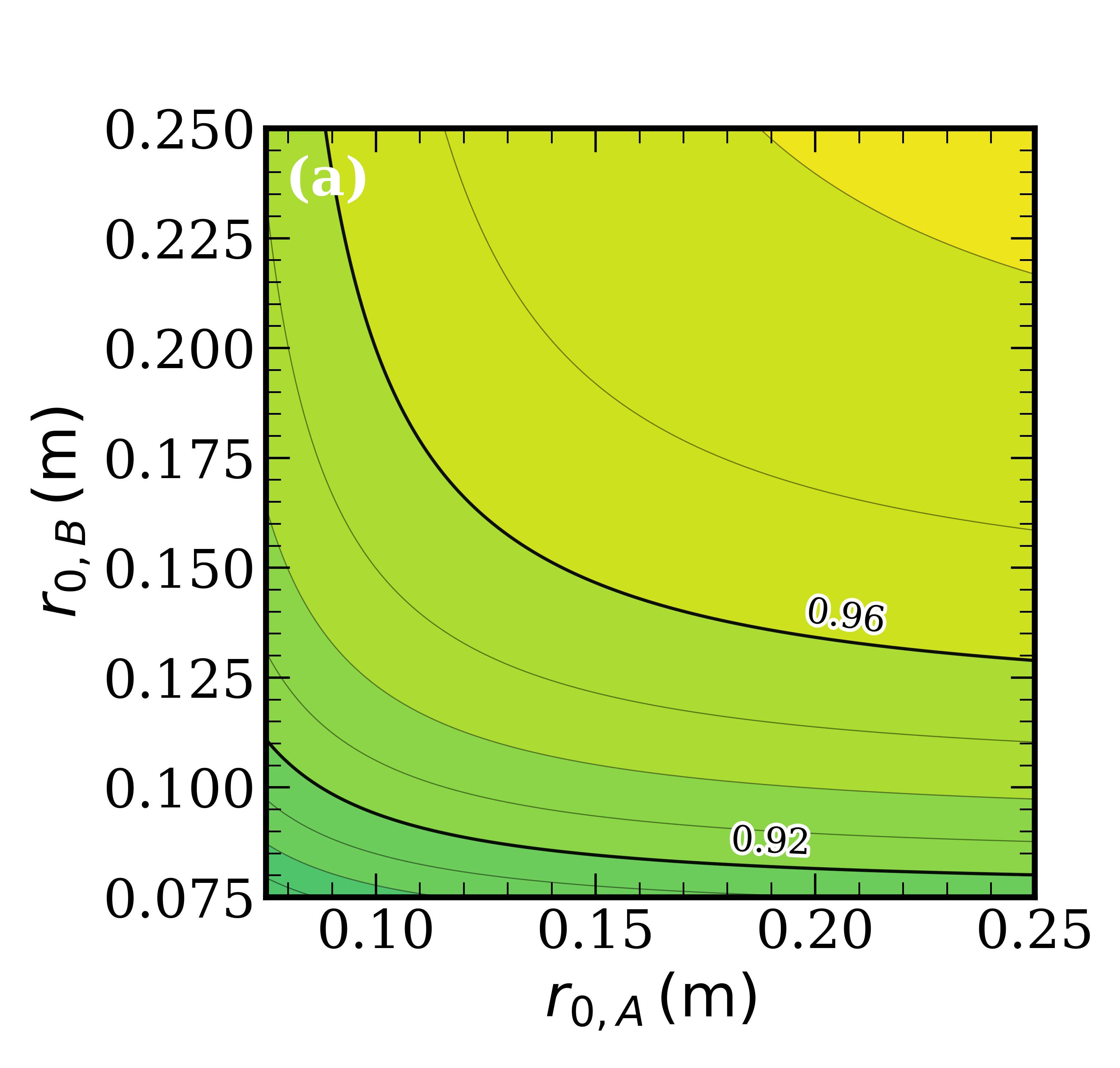}
    \end{minipage}\hfill
    \begin{minipage}[t]{0.49\linewidth}
        \centering
        \includegraphics[height=6.8cm, keepaspectratio]{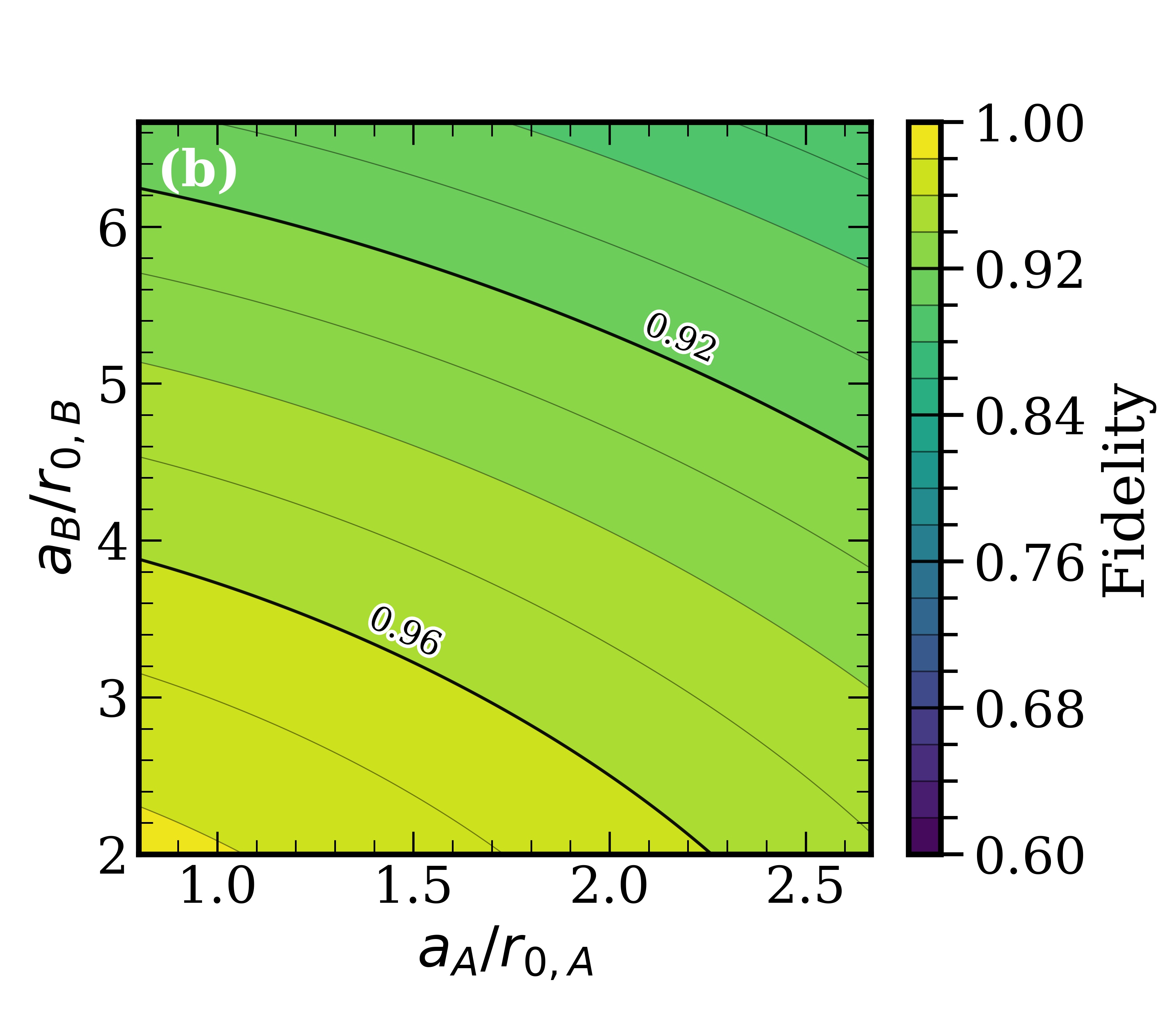}
    \end{minipage}
    \caption{Scenario II: asymmetric turbulence parameters $\kappa_A=0.01$ and $\kappa_B=0.005$. Panel (a) shows the fidelity as a function of the Fried parameters for unequal apertures for $a_A=0.2\,\mathrm{m}$ and $a_B=0.5\,\mathrm{m}$. Panel (b) gives the equivalent dimensionless representation in terms of $a_A/r_{0,A}$ and $a_B/r_{0,B}$. The states coefficients are: $\alpha=\frac{1}{\sqrt{2}}$, $ \beta = \sqrt{1 - {\lvert \alpha  \rvert}^2}$, $\gamma = \frac{i}{\sqrt{2}}$, $\delta = \sqrt{1- {\lvert \gamma \rvert}^2}$.}
    \label{fig:SCENARIO_B}
\end{figure*}

\begin{figure*}[t]
    \centering
    \begin{minipage}[t]{0.49\linewidth}
        \centering
        \includegraphics[height=6.8cm, keepaspectratio]{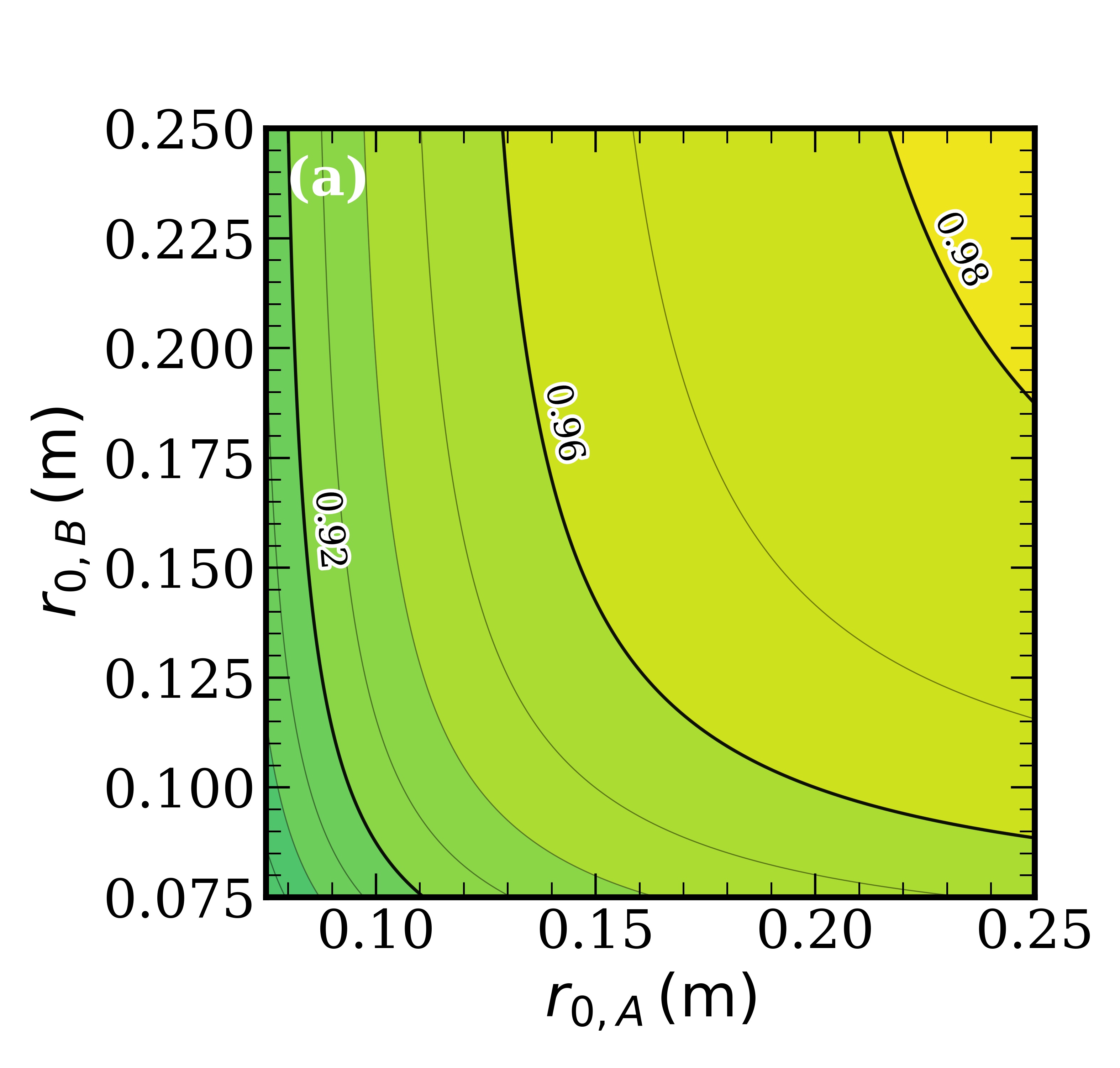}
    \end{minipage}\hfill
    \begin{minipage}[t]{0.49\linewidth}
        \centering
\includegraphics[height=6.8cm, keepaspectratio]{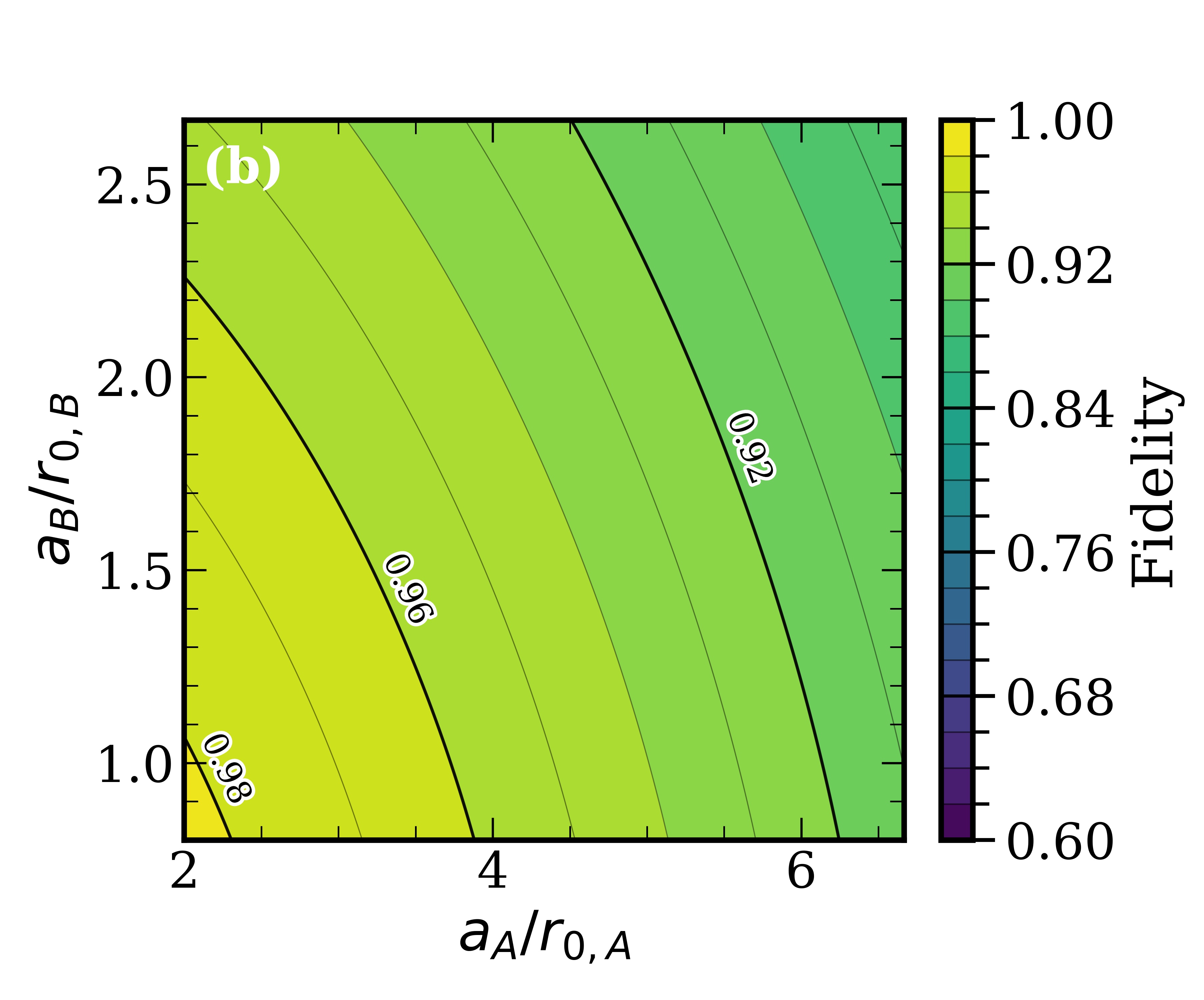}
    \end{minipage}
    \caption{Scenario III: exchanged turbulence strengths, $\kappa_A=0.005$ and $\kappa_B=0.01$. Panel (a) shows the fidelity as a function of the Fried parameters for unequal apertures $a_A=0.5\,\mathrm{m}$ and $a_B=0.2\,\mathrm{m}$. Panel (b) presents the same dependence in terms of the dimensionless ratios $a_A/r_{0,A}$ and $a_B/r_{0,B}$ The states coefficients are: $\alpha=\frac{1}{\sqrt{2}}$, $ \beta = \sqrt{1 - {\lvert \alpha  \rvert}^2}$, $\gamma = \frac{i}{\sqrt{2}}$, $\delta = \sqrt{1- {\lvert \gamma \rvert}^2}$.}
    \label{fig:SCENARIO_C}
\end{figure*}

\begin{figure*}[t]
    \centering
    \begin{minipage}[t]{0.49\linewidth}
        \centering
        \includegraphics[height=6.8cm, keepaspectratio]{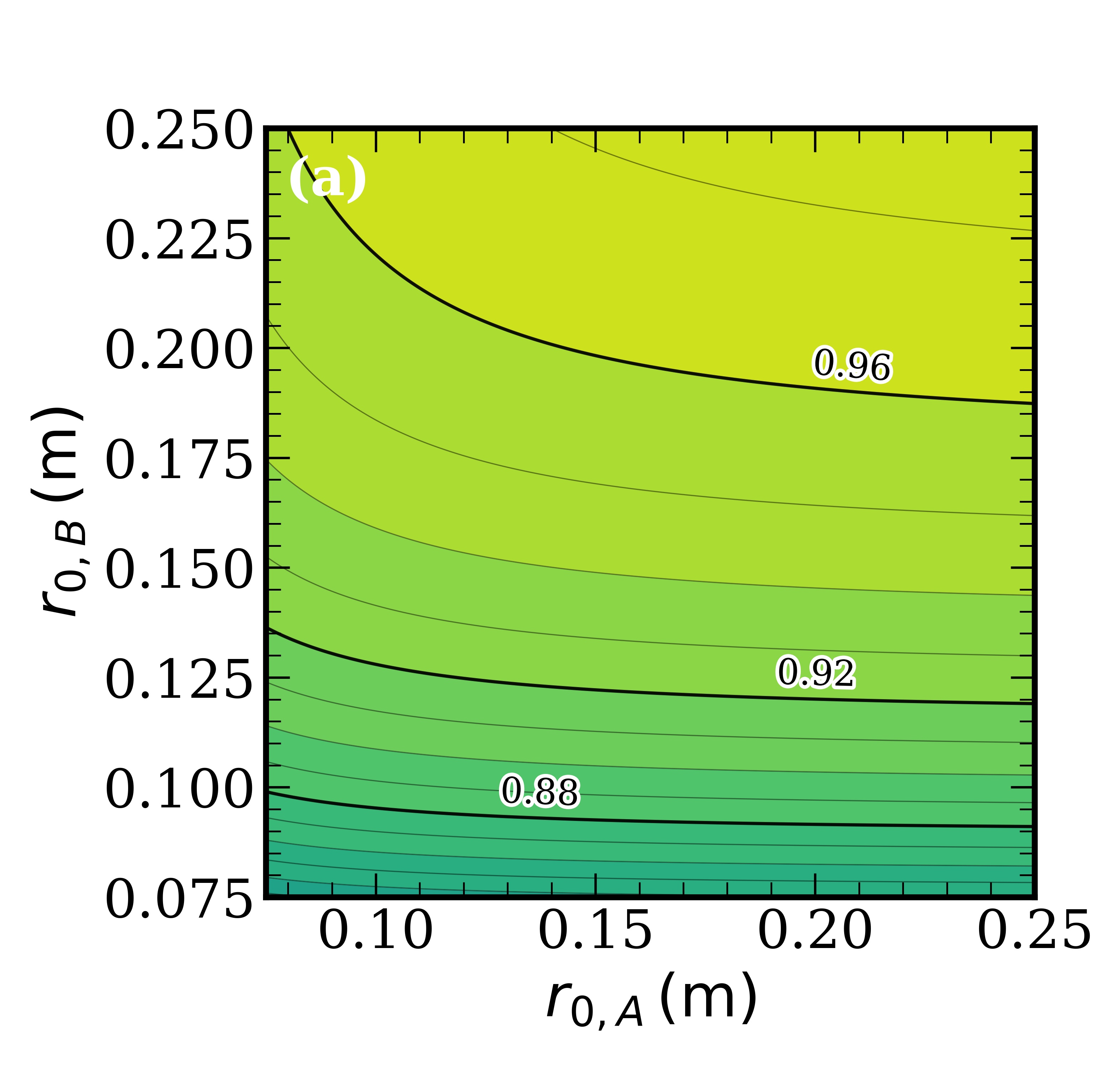}
    \end{minipage}\hfill
    \begin{minipage}[t]{0.49\linewidth}
        \centering
        \includegraphics[height=6.8cm, keepaspectratio]{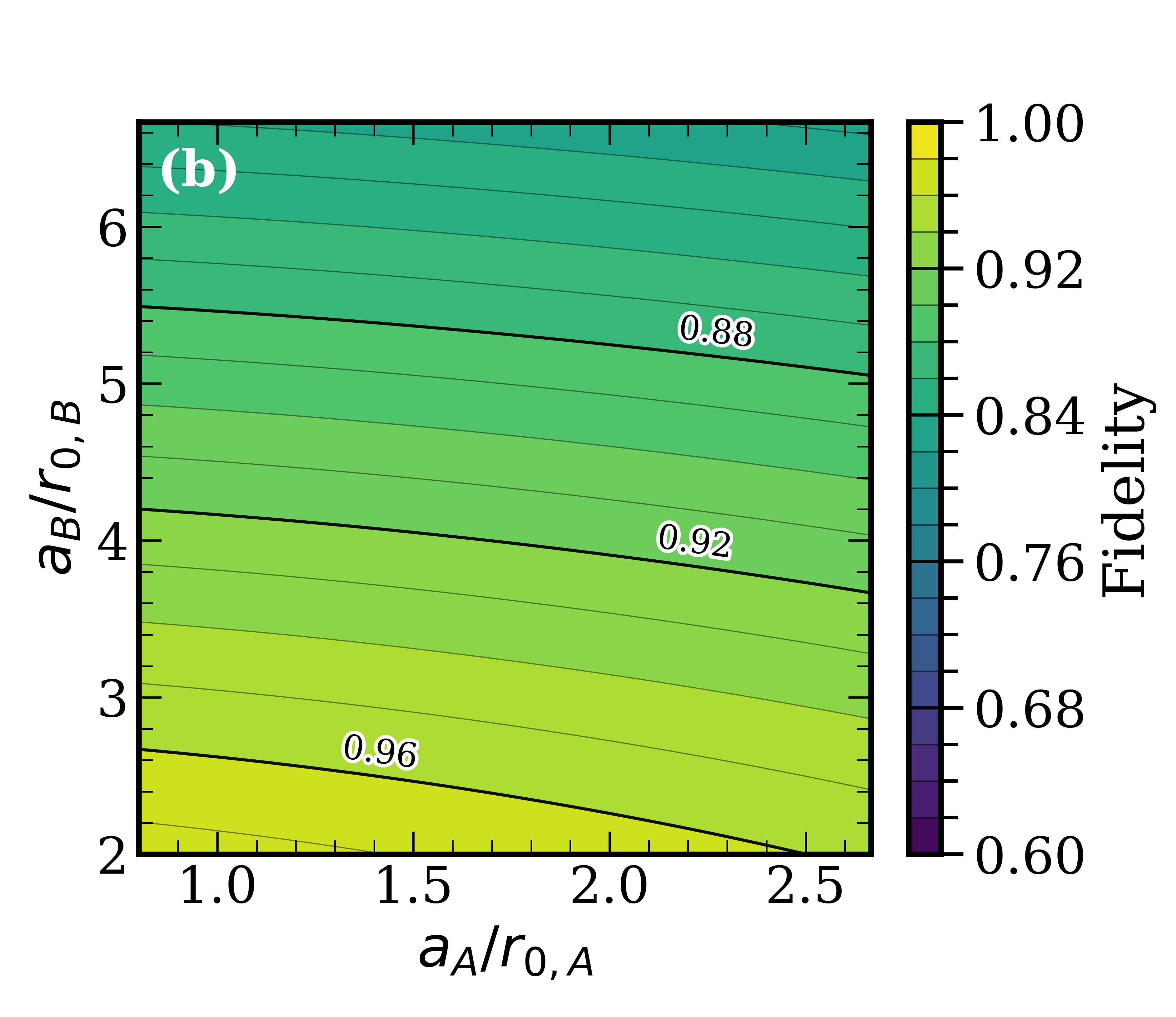}
    \end{minipage}
    \caption{Scenario IV: asymmetric turbulence $\kappa_A=0.005$ and $\kappa_B=0.01$. Panel (a) shows the fidelity as a function of the Fried parameters $r_{0,A}$ and $r_{0,B}$ for unequal apertures $a_A=0.2\,\mathrm{m}$ and $a_B=0.5\,\mathrm{m}$. Panel (b) presents the same dependence in terms of the dimensionless ratios $a_A/r_{0,A}$ and $a_B/r_{0,B}$. This scenario mainly tests the asymmetry between channels A and B. The states coefficients are: $\alpha=\frac{1}{\sqrt{2}}$, $ \beta = \sqrt{1 - {\lvert \alpha  \rvert}^2}$, $\gamma = \frac{i}{\sqrt{2}}$, $\delta = \sqrt{1- {\lvert \gamma \rvert}^2}$.}
    \label{fig:SCENARIO_D}
\end{figure*}

\begin{figure*}[t]
    \centering
    \begin{minipage}[t]{0.49\linewidth}
        \centering     \includegraphics[height=6.8cm, keepaspectratio]{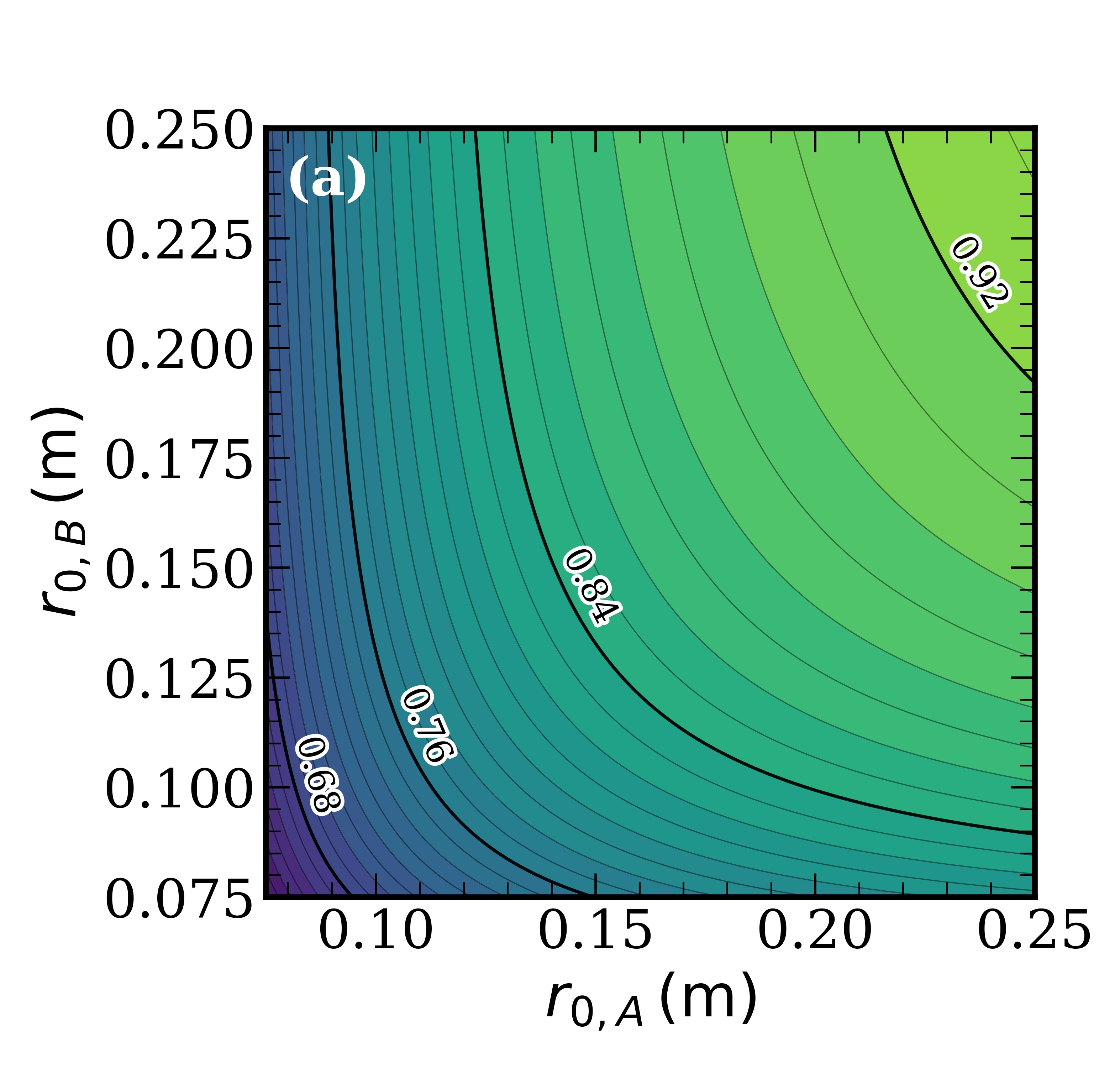}
    \end{minipage}\hfill
    \begin{minipage}[t]{0.49\linewidth}
        \centering
\includegraphics[height=6.8cm, keepaspectratio]{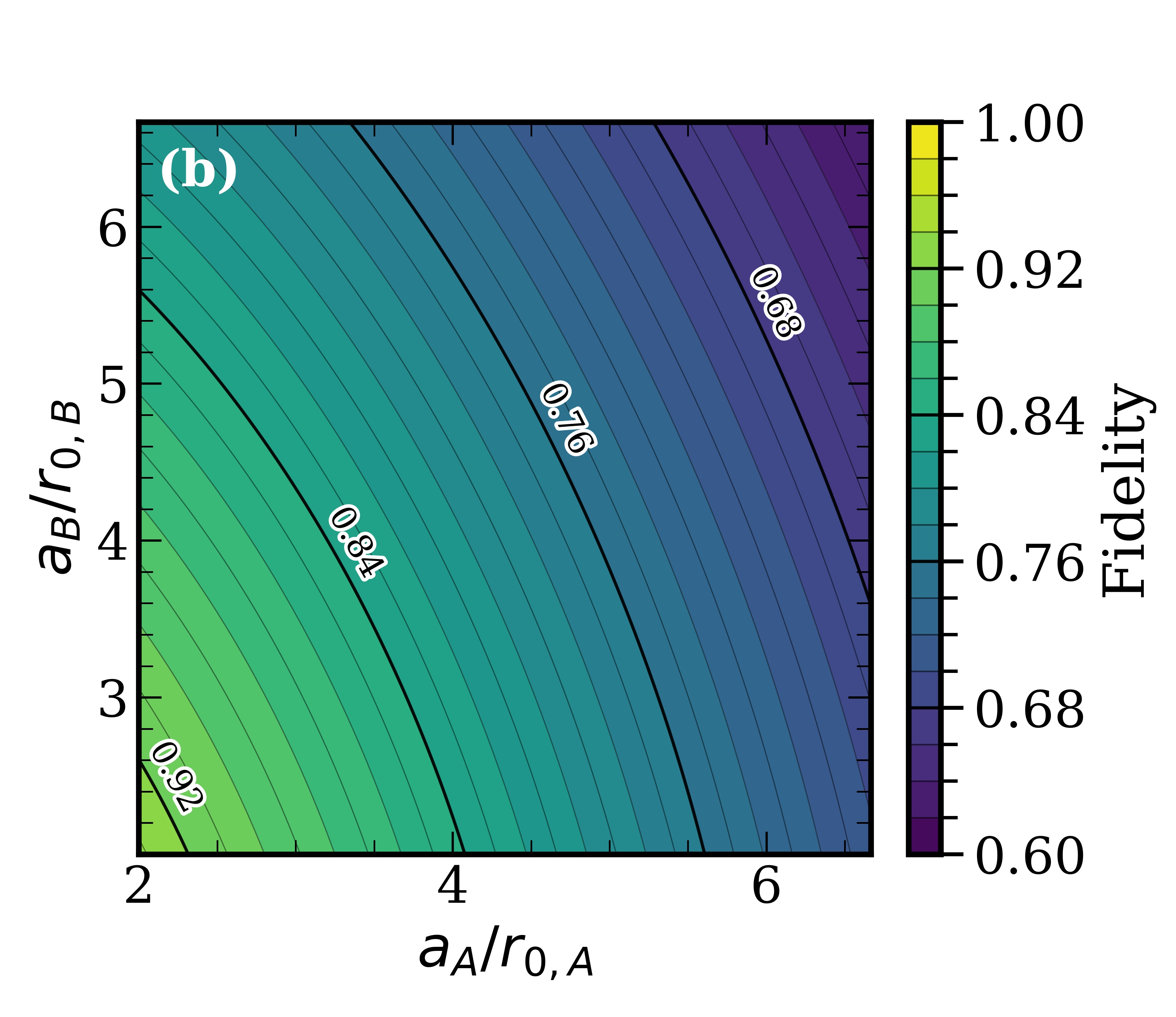}
    \end{minipage}
    \caption{Scenario V: stronger-turbulence stress test with increased degradation parameters $\kappa_A=0.02$ and $\kappa_B=0.01$. Panel (a) shows the fidelity as a function of the Fried parameters $r_{0,A}$ and $r_{0,B}$ for equal apertures $a_A=a_B=0.5\,\mathrm{m}$. Panel (b) presents the same dependence in terms of the dimensionless ratios $a_A/r_{0,A}$ and $a_B/r_{0,B}$. The states coefficients are: $\alpha=\frac{1}{\sqrt{2}}$, $ \beta = \sqrt{1 - {\lvert \alpha  \rvert}^2}$, $\gamma = \frac{i}{\sqrt{2}}$, $\delta = \sqrt{1- {\lvert \gamma \rvert}^2}$.}
    \label{fig:SCENARIO_E}
\end{figure*}

In this subsection, five scenarios were considered in order to analyze the influence of atmospheric coherence and receiver-aperture configurations on the state fidelity of the QGT protocol implementing a remote CNOT gate. In this part of the analysis, the input states of the circuit qubits remain fixed, so that the observed changes in fidelity can be attributed to the parameters of the optical links. The considered scenarios differ in the receiving aperture radii, denoted by $a_A$ and $a_B$, as well as in the turbulence parameters $\kappa_A$ and $\kappa_B$, assigned to channels $A$ and $B$. The effective depolarization is modeled according to Eq.~(\ref{eq:pdep_general}) with $\epsilon = 5/3$. They include the case of equal apertures, configurations with different apertures on both sides of the link, exchanged channel parameters, and a test case corresponding to stronger turbulence-induced degradation.

The simulation results for these scenarios are presented in Figs.~\ref{fig:SCENARIO_A}-\ref{fig:SCENARIO_E}. In each figure, panel (a) shows the fidelity as a function of the Fried parameters $r_{0,A}$ and $r_{0,B}$, which represent the effective atmospheric coherence conditions in both optical channels. Panel (b) presents the same dependence as a function of the dimensionless aperture-to-coherence-length ratios, $a_A/r_{0,A}$ and $a_B/r_{0,B}$. This representation makes it possible to assess how the quality of the remote quantum operation depends both on the atmospheric coherence described by the Fried parameters and on the relative size of the receiver apertures with respect to the Fried coherence length.

The same linear color normalization, $F\in[0.60,1.00]$, is used for all fidelity maps. The lower limit, $F=0.60$, corresponds to the lowest values of fidelity Eq.~(\ref{fidelitydef}) encountered across the considered set of scenarios, whereas $F=1.00$ represents the maximum possible fidelity. Consequently, a given color corresponds to the same fidelity value in every map in Section~\ref{remotecnotgate}, allowing direct quantitative comparison between different scenarios and panels.
The filled color contours are defined with a spacing of $\Delta F=0.02$, while additional thin contour lines are shown at intervals of $\Delta F=0.01$ to reveal finer local variations without altering the common color normalization. Selected reference fidelity levels are additionally emphasized by thicker black contours and labeled directly on the maps. The same set of emphasized fidelity levels is used for the two panels, (a) and (b), belonging to a given scenario, thereby facilitating direct comparison between the corresponding configurations. Only a limited number of reference contours are labeled in order to preserve visual clarity; intermediate fidelity values can be inferred from the thinner contour lines and from the common color scale.

The obtained results show an increase in the fidelity of the QGT protocol with increasing Fried parameters $r_{0,A}$ and $r_{0,B}$ in both optical channels. The lowest fidelity values occur for small values of $r_{0,A}$ and $r_{0,B}$, whereas for larger values of the Fried parameters the fidelity systematically increases and reaches values close to unity.

The comparison of Scenarios I-IV shows that changes in the aperture sizes $a_A$ and $a_B$, as well as in the parameters $\kappa_A$ and $\kappa_B$, affect the fidelity results. In scenarios with unequal apertures, an asymmetry with respect to channels A and B is visible, depending on which side is assigned the larger aperture and the larger value of the parameter $\kappa$. The obtained surfaces differ in the range of fidelity values and in the slope of the changes with respect to the parameters $r_{0,A}$ and $r_{0,B}$.

The lowest fidelity values among the analyzed cases are obtained in Scenario V in Fig.~\ref{fig:SCENARIO_E}, where increased values of the parameters $\kappa_A$ and $\kappa_B$ are used. In this case, the fidelity range is shifted towards lower values compared with Scenarios I-IV. For the remaining scenarios, the minimum fidelity values remain higher, whereas in the region of large Fried parameters higher fidelity values are observed for all configurations.

\subsection{QGT Fidelity for Different Circuit-Qubit States}\label{sec:DifferentCircuit}

\begin{figure*}[t]
    \centering
    \begin{minipage}[t]{0.49\linewidth}
        \centering
    \includegraphics[height=6.8cm, keepaspectratio]{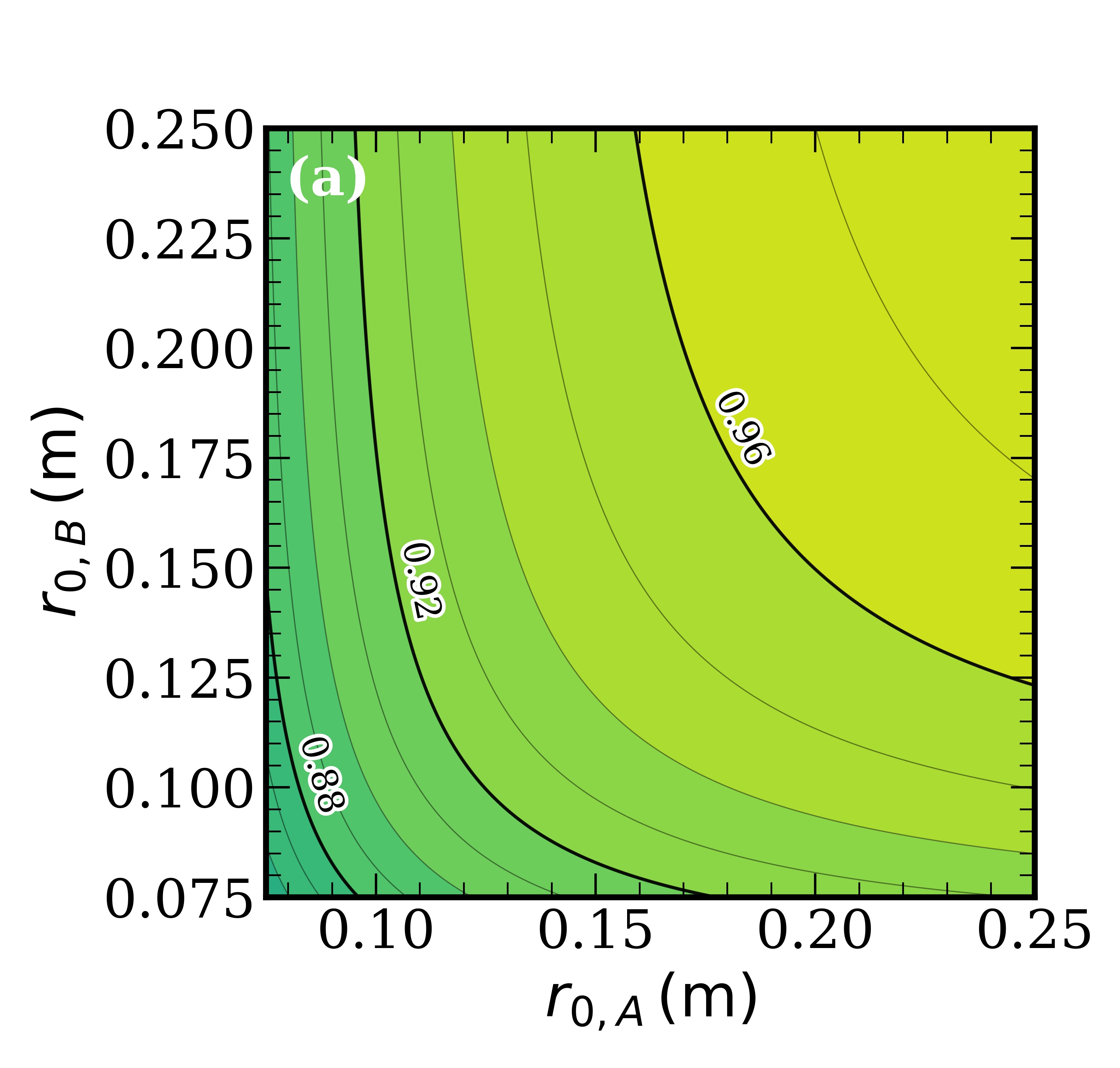}
    \end{minipage}\hfill
    \begin{minipage}[t]{0.49\linewidth}
        \centering
        \includegraphics[height=6.8cm, keepaspectratio]{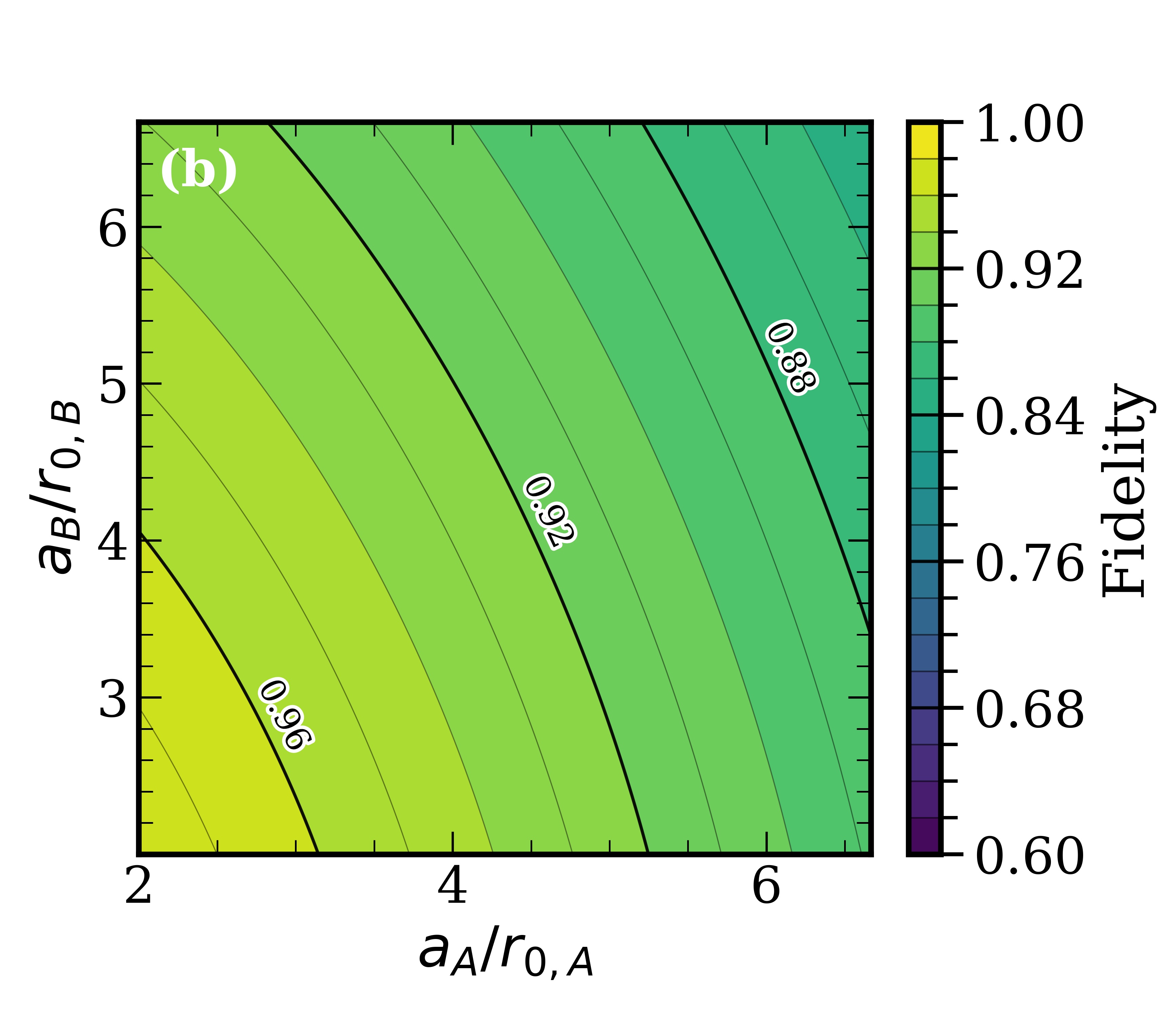}
    \end{minipage}
    \caption{Scenario VI: turbulence parameters $\kappa_A=0.01$ and $\kappa_B=0.005$. Panel (a) shows the fidelity as a function of the Fried parameters $r_{0,A}$ and $r_{0,B}$ for equal apertures $a_A=a_B=0.5\,\mathrm{m}$. Panel (b) presents the same quantity in terms of the dimensionless ratios $a_A/r_{0,A}$ and $a_B/r_{0,B}$. The states coefficients are: $\alpha=\frac{1}{\sqrt{2}}$, $ \beta = \sqrt{1 - {\lvert \alpha  \rvert}^2}$, $\gamma = \frac{1}{\sqrt{2}}$, $\delta = \sqrt{1- {\lvert \gamma \rvert}^2}$.}
    \label{fig:SCENARIO_F1}
\end{figure*}

\begin{figure*}[t]
    \centering
    \begin{minipage}[t]{0.49\linewidth}
        \centering
        \includegraphics[height=6.8cm, keepaspectratio]{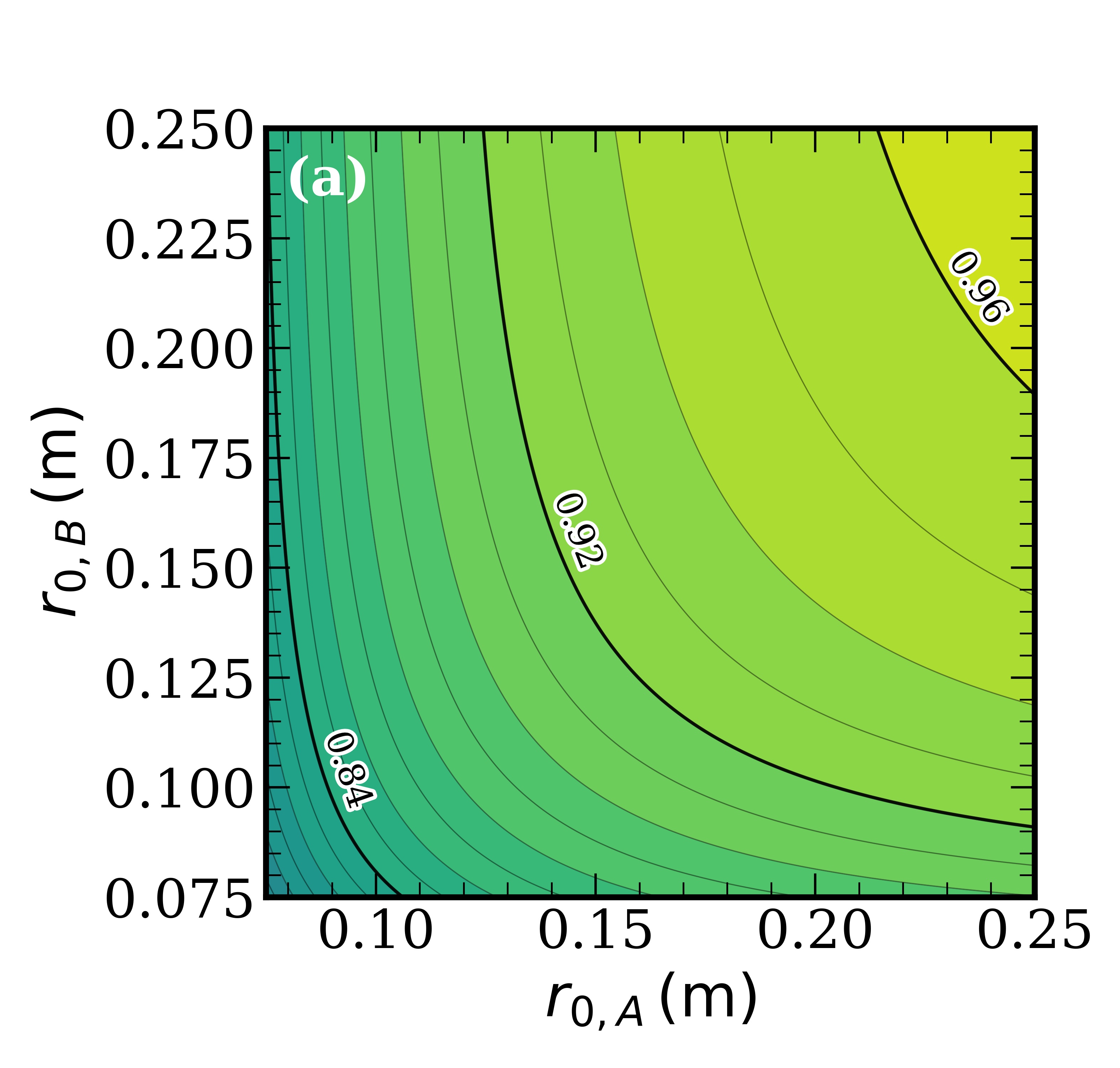}
    \end{minipage}\hfill
    \begin{minipage}[t]{0.49\linewidth}
        \centering
        \includegraphics[height=6.8cm, keepaspectratio]{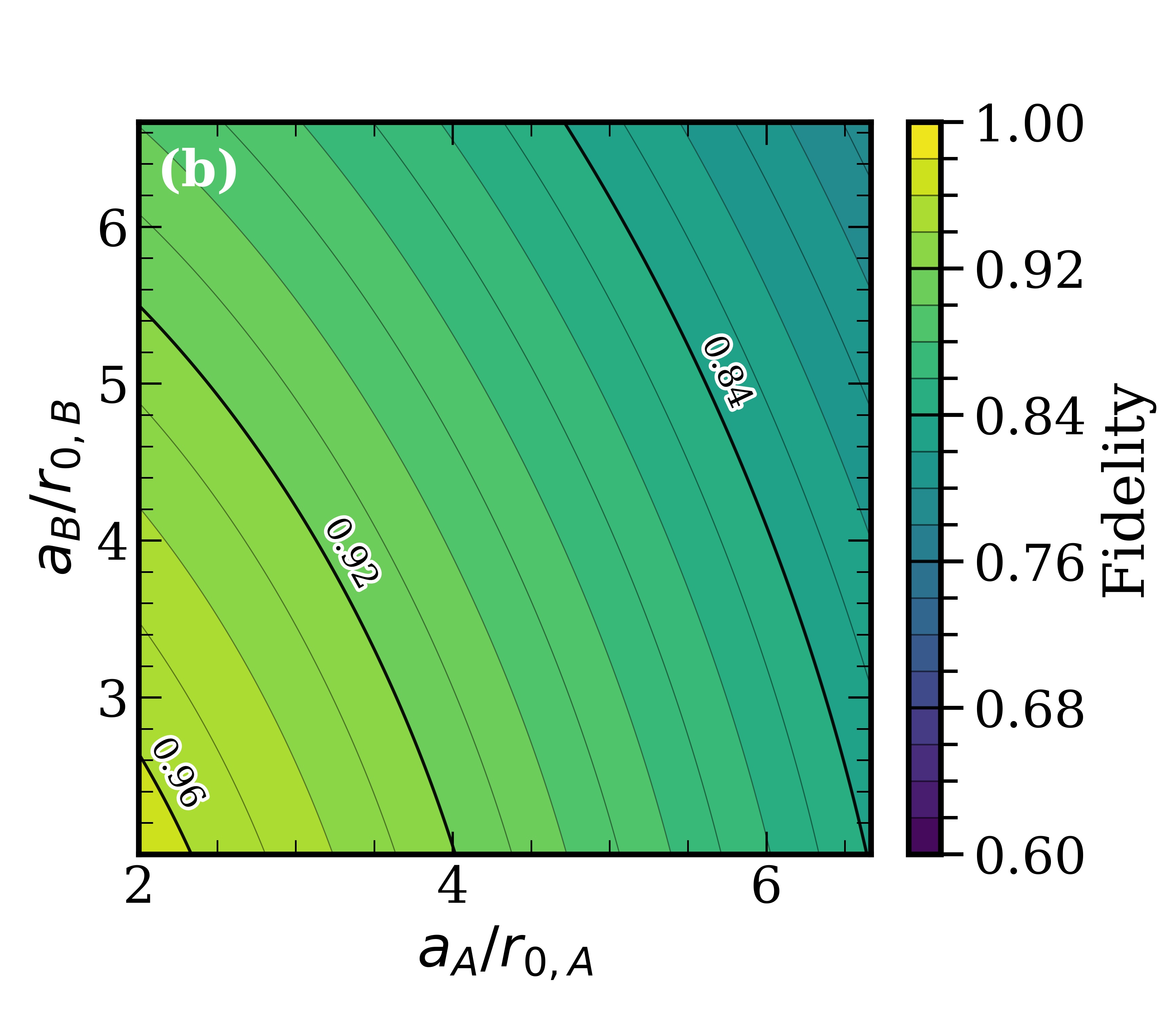}
    \end{minipage}
    \caption{Scenario VII: turbulence parameters $\kappa_A=0.01$ and $\kappa_B=0.005$. Panel (a) shows the fidelity as a function of the Fried parameters $r_{0,A}$ and $r_{0,B}$ for equal apertures $a_A=a_B=0.5\,\mathrm{m}$. Panel (b) presents the same quantity in terms of the dimensionless ratios $a_A/r_{0,A}$ and $a_B/r_{0,B}$. In this case we use asymmetric input-state amplitudes for the same baseline channel configuration. The states coefficients are: $\alpha=\frac{1}{\sqrt{1.5}}$, $ \beta = \sqrt{1 - {\lvert \alpha  \rvert}^2}$, $\gamma = \frac{i}{\sqrt{2}}$, $\delta = \sqrt{1- {\lvert \gamma \rvert}^2}$.}
    \label{fig:SCENARIO_F2}
\end{figure*}

\begin{figure*}[t]
    \centering
    \begin{minipage}[t]{0.49\linewidth}
        \centering
        \includegraphics[height=6.8cm, keepaspectratio]{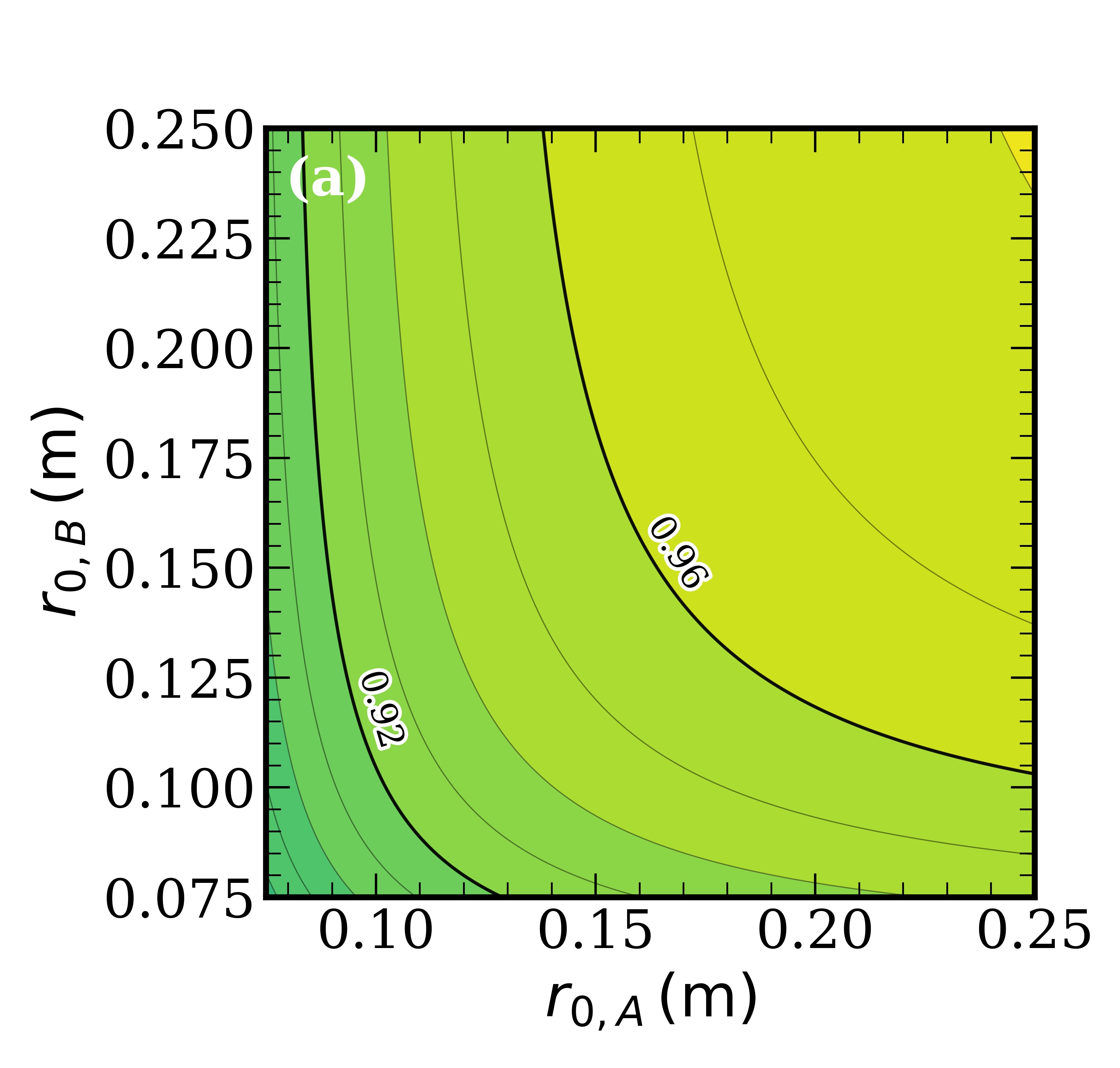}
    \end{minipage}\hfill
    \begin{minipage}[t]{0.49\linewidth}
        \centering
        \includegraphics[height=6.8cm, keepaspectratio]{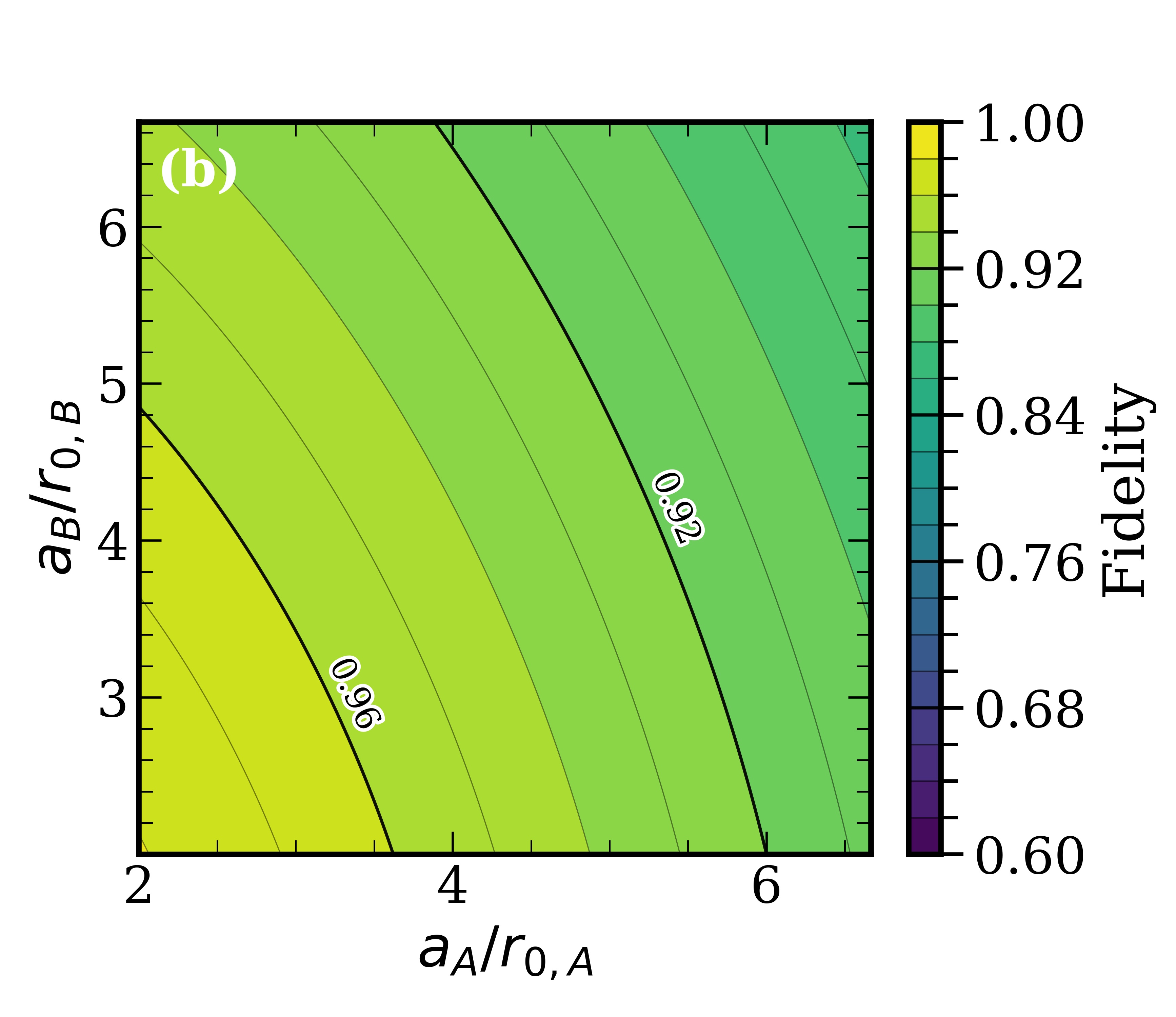}
    \end{minipage}
    \caption{Scenario VIII: turbulence parameters $\kappa_A=0.01$ and $\kappa_B=0.005$. Panel (a) shows the fidelity as a function of the Fried parameters $r_{0,A}$ and $r_{0,B}$ for equal apertures $a_A=a_B=0.5\,\mathrm{m}$. Panel (b) presents the same dependence in terms of the dimensionless ratios $a_A/r_{0,A}$ and $a_B/r_{0,B}$. The states coefficients are: $\alpha=\sqrt{0.85}$, $ \beta = \sqrt{1 - {\lvert \alpha  \rvert}^2}$, $\gamma = \frac{1}{\sqrt{2}}$, $\delta = \sqrt{1- {\lvert \gamma \rvert}^2}$.}
    \label{fig:SCENARIO_F3}
\end{figure*}

In this subsection, we investigate the influence of the input states of the circuit qubits on the output-state fidelity of the QGT protocol. In contrast to the previous subsection, the receiver-aperture configuration and the turbulence-related parameters are kept fixed. In all analyzed cases, equal apertures $a_A = a_B = 0.5,\mathrm{m}$ and the turbulence parameters $\kappa_A = 0.01$ and $\kappa_B = 0.005$ are assumed. The scenarios differ only in the input-state coefficients $\alpha$, $\beta$, $\gamma$, and $\delta$, which define the initial states of the circuit qubits $A$ and $B$.

The simulation results for these cases are presented in Figs.~\ref{fig:SCENARIO_F1}--\ref{fig:SCENARIO_F3}. In each figure, panel (a) shows the fidelity as a function of the Fried parameters $r_{0,A}$ and $r_{0,B}$, whereas panel (b) presents the same dependence as a function of the dimensionless ratios $a_A/r_{0,A}$ and $a_B/r_{0,B}$. This representation makes it possible to compare how different input circuit-qubit states affect the fidelity of the remote CNOT operation under the same receiver-aperture configuration and turbulence-related parameters.

The obtained results show that, for all analyzed input states of the circuit qubits, the fidelity of the QGT protocol increases with increasing Fried parameters $r_{0,A}$ and $r_{0,B}$. The lowest fidelity values are observed in the region of small Fried-parameter values, whereas the improvement of atmospheric coherence conditions in both channels leads to a systematic increase in the fidelity of the remote CNOT operation.

The comparison of Scenarios VI--VIII shows that changing the input-state coefficients $\alpha$, $\beta$, $\gamma$, and $\delta$ affects the obtained fidelity results, even though the receiver-aperture configuration and the turbulence-related parameters remain the same. The differences between the scenarios mainly concern the level of the obtained fidelity, whereas the general character of the dependence on the parameters $r_{0,A}$ and $r_{0,B}$ remains similar.

This input-state dependence reflects how noise in the shared entangled resource propagates through the local operations of the QGT protocol. Within each QPU, the circuit qubit interacts with a network qubit belonging to the distributed entangled pair. The resulting errors can affect circuit states differently depending on their orientation relative to the bases selected by the protocol's gates and measurements. Consequently, the same noisy entangled resource can yield different QGT fidelities for different circuit input states.

In all three cases from Figs.~\ref{fig:SCENARIO_F1}--\ref{fig:SCENARIO_F3}, the dimensionless representation confirms that an increase in the ratios $a_A/r_{0,A}$ and $a_B/r_{0,B}$ is associated with a decrease in fidelity. This means that, for fixed channel parameters, different input states of the circuit qubits may change the final fidelity value, while the dominant trend dependent on the effective atmospheric conditions remains preserved.

\section{Discussion and Conclusions}\label{conclusions}

The results presented in Sections~\ref{remotecnotgateres}-\ref{sec:DifferentCircuit} indicate that the quality of the QGT protocol is directly related to the effective degradation of the entangled resource used to implement the remote CNOT gate. An increase in the Fried parameters in both optical channels leads to improved fidelity, since larger values of $r_0$ correspond to better atmospheric coherence and lower effective depolarization. Conversely, an increase in the ratio between the aperture size and the Fried coherence length reduces the fidelity, which is consistent with the adopted model, where larger values of $a/r_0$ correspond to stronger effective degradation because the receiver samples a larger region relative to the Fried coherence length.

The comparison of scenarios with different apertures and different values of the parameters $\kappa_A$ and $\kappa_B$ shows that the final fidelity does not depend on a single channel parameter, but rather on their combined effect. In particular, the combination of aperture size, Fried parameter, and the effective susceptibility of the system to polarization mixing plays an important role. The results obtained for the scenario with increased values of $\kappa_A$ and $\kappa_B$ confirm that stronger effective channel degradation can significantly limit the quality of the remote CNOT operation.

The analysis of different input states of the circuit qubits shows that changing the coefficients $\alpha$, $\beta$, $\gamma$, and $\delta$ may affect the obtained fidelity values, while the general trend of the dependence on atmospheric parameters remains preserved. This indicates that, within the analyzed model, one of the dominant factors limiting the protocol performance is the quality of the distributed entangled resource, which depends on the optical-channel parameters. In practice, this points to the need for simultaneous optimization of propagation conditions, receiver geometry, and polarization stability of the optical system in future satellite-based QGT implementations.

The significance of these results lies in showing that QGT can be treated not only as an abstract primitive of distributed quantum computing, but also as a channel-sensitive protocol whose performance can be directly connected to measurable properties of a satellite optical link. In particular, the fidelity of the remotely implemented CNOT gate provides a compact operational figure of merit: it translates atmospheric turbulence, receiver aperture effects, and polarization degradation into the quality of a distributed quantum operation. This is important because satellite-based quantum networks are often discussed primarily in terms of entanglement distribution or quantum key distribution, whereas distributed quantum computing requires a more demanding assessment at the level of executable nonlocal gates.

From this perspective, the present analysis provides a bridge between physical-layer channel modeling and higher-level quantum information processing. The obtained results indicate which atmospheric and optical parameters are most relevant for preserving the usefulness of satellite-distributed entanglement as a computational resource. In particular, they show that even if entanglement distribution is possible in principle, its quality may not be sufficient for high-fidelity quantum gate teleportation unless depolarization effects are kept under control. Therefore, the results can help identify operational regimes in which satellite links may support not only communication-oriented protocols, but also elementary distributed computing tasks.

Recent experimental progress shows that QGT is becoming a realistic operational primitive for distributed quantum computing, but practical implementations remain at an early stage. In particular, Main \textit{et al.} demonstrated deterministic quantum gate teleportation between two photonically interconnected trapped-ion modules separated by approximately $\sim2$ m, achieving a teleported two-qubit gate fidelity of ($86\%$) and using this capability to execute elementary distributed quantum circuits~\cite{main2025distributed}. This result is an important proof of principle, but it also illustrates that current demonstrations are still limited to short-distance laboratory-scale links and carefully controlled experimental conditions. Therefore, extending QGT to satellite-enabled or global-scale architectures requires detailed numerical modeling to determine the operational limits, expected fidelity ranges, and sensitivity to atmospheric and optical-channel parameters. Such modeling is necessary to identify whether satellite-distributed entanglement can support gate-level operations with sufficient quality for large-scale distributed quantum computing.

The study also highlights the importance of receiver design and channel characterization for future implementations of satellite-enabled distributed quantum computing. Since the fidelity depends on the combined influence of the Fried parameter, aperture size, and effective depolarization parameters, the optimization of a satellite quantum link cannot be reduced to maximizing received optical power alone. Instead, the preservation of polarization correlations and the stability of the entangled resource must be considered jointly with link-budget constraints. This conclusion is particularly relevant for the design of future quantum network architectures, where satellite nodes may be expected to distribute entanglement between remote processors and enable gate-level operations between them.

It should be noted that the present analysis focuses on the effective depolarizing contribution to the atmospheric quantum channel. Other imperfections of satellite optical links, such as propagation losses, background counts, and Poisson noise, are not explicitly modeled in the QGT fidelity calculations. These effects may influence the probability of successfully distributing and detecting the entangled resource, as well as the rate describing the accessible speed of quantum computations. Such effects should therefore be included in future extensions of the model. A more complete analysis combining link-budget effects with polarization noise would allow a more accurate assessment of satellite-based QGT under realistic operational conditions.

\section*{Funding}
J.J.B. was supported by the National Science Centre of Poland (Grant No. 2020/39/O/ST2/00137).

\section*{Data Availability}
No datasets were generated or analyzed during the current study.

\section*{Conflict of Interest}
The author declares that there is no conflict of interest.

\end{document}